\documentclass[]{aa} % for a referee version
\usepackage{graphicx}
\usepackage{txfonts}
\usepackage{epsfig}
\begin{document}
   \title{An $I$-band calibration of the SBF method at blue colours}

   %\subtitle{$I$-band SBF calibration}

   \author{S. Mieske
          \inst{1}
          \and
          M. Hilker\inst{2}\and L. Infante\inst{3}
          }

   \offprints{S. Mieske}

   \institute{European Southern Observatory, Karl-Schwarzschild-Strasse 2, 85748 Garching bei M\"unchen, Germany\\
              \email{smieske@eso.org}
         \and
             Argelander Institut f\"ur Astronomie - Abteilung Sternwarte,
Auf dem H\"ugel 71, 53121 Bonn, Germany\\
             \email{mhilker@astro.uni-bonn.de}
         \and
             Departamento de Astronom\'{\i}a y Astrof\'{\i}sica, Pontificia
Universidad Cat\'olica de Chile, Casilla 306, Santiago 22, Chile\\
             \email{linfante@astro.puc.cl}
             }

   \date{}
\titlerunning{SBF calibration at blue colours}

\authorrunning{S.~Mieske et al.}
  \abstract
   {The surface brightness fluctuation (SBF) method is a powerful tool to derive distances to galaxies for which single stars cannot be resolved. Up to now, the method has been calibrated mainly at red colours due to the intrinsic faintness of blue early-type galaxies.}
   {In this paper we address the $I$-band calibration of the SBF method at blue colours, the regime of dwarf elliptical galaxies (dEs).}
   {We present deep and wide-field $VI$ photometry of the central Fornax cluster obtained at Las Campanas Observatory. With these data we perform an SBF analysis of 25 dEs in the range $-16.5<M_V<-11.2$ mag, $0.8<(V-I)_0<1.10$ mag. Our colour calibration is accurate to better than 2\%. For the calibration analysis we exclude eight dEs whose SBF measurement was affected by poor seeing (FWHM $\simeq$ 1$''$).}
   {Our SBF data are inconsistent at the 3$\sigma$ level with a colour independent absolute SBF magnitude $\overline{M}_I$, presenting a problem for SBF models that predict such a ``flat'' relation. There is a weak indication in our data (1.8$\sigma$) to favour a two-branch calibration over a one-branch calibration with broad scatter in the $\overline{M}_I - (V-I)$ plane. We obtain the following one-branch empirical SBF calibration:  $\overline{M}_I=-2.13 (\pm 0.17) + 2.44 (\pm 1.94) \times [(V-I)_{\rm 0} - 1.00]\;{\rm mag}$. We deduce a 0.34 $\pm$ 0.14 mag cosmic scatter of $\overline{M}_I$, which is significantly larger than found at redder colours. This is in agreement with those theoretical SBF models that predict $\overline{M}_I$ to be more sensitive to age-metallicity variations in the blue than in the red. We find evidence that the fainter galaxies in our samples contain younger and more metal-rich stellar populations than the brighter ones.
The application of our empirical calibration to published SBF measurements of Hydra and Centaurus cluster dEs leaves the distances to both clusters unchanged.}
   {}


   \maketitle
%
%________________________________________________________________

\section{Introduction}
\subsection{The principle of SBF}
The measurement of surface brightness fluctuations (SBF) in a galaxy image was first suggested by Tonry \& Schneider~(\cite{Tonry88}) as a useful tool to measure distances to galaxies which are not resolved into single stars (see also Jacoby et al.~\cite{Jacoby92} for a review). The SBF method is based on two facts:\\
A) the number of stars contained in each resolution element of a galaxy image is {\it finite}.\\
B) the number of stars contained in each resolution element is on average {\it constant} along one isophote.\\
From these two facts, a third fact is deduced: the number of stars per resolution element along one isophote has a statistical fluctuation equal to the square root of the average star number. The amplitude of those surface brightness fluctuations normalised to the underlying mean galaxy luminosity is inversely proportional to distance and can therefore be used as a distance indicator. This normalised amplitude converted to magnitude is in the framework of SBF investigations referred to as apparent fluctuation magnitude $\overline{m}$. It corresponds to the luminosity weighted mean luminosity of stars in the observed stellar population (see e.g. Tonry \& Schneider~\cite{Tonry88} and Liu et al.~\cite{Liu00}).\\
The highest luminosity stars in an old stellar population are the red giants. Consequently, the average number of stars per unit surface brightness is lowest in the colour regime of red giants. SBF measurements have therefore mainly been performed in red optical pass-bands (e.g. Tonry \& Schneider~\cite{Tonry88}, Bothun et al.~\cite{Bothun91}, Tonry et al.~\cite{Tonry97}, Jerjen et al. \cite{Jerjen04}, Mei et al.~\cite{Mei05}) and in the near infrared (e.g. Luppino \& Tonry \cite{Luppin93}, Pahre \& Mould \cite{Pahre94}, Liu \& Graham \cite{Liu01}, Liu et al.~\cite{Liu02}, Jensen et al.~\cite{Jensen03}), both from the ground and with HST.
\subsection{SBF calibrations}
Of course, any distance determination method needs to be calibrated in order to derive absolute distances. In their pioneering work, Tonry et al. (\cite{Tonry97}, \cite{Tonry01}) have carried out 
an extensive survey to measure the apparent $I$-band SBF magnitude $\overline{m}_I$ for several hundred bright ellipticals and bulges of spiral galaxies in 
22 nearby galaxy groups and clusters within 40 Mpc. 
Using distances derived from Cepheids or SN~Ia in the respective galaxies or the clusters they are associated with, 
they obtained a purely empirical relation between the absolute I-band fluctuation magnitude $\overline{M}_I$ 
and the dereddened colour $(V-I)_{\rm 0}$, determined in the range $1.05<(V-I)_{\rm 0}<1.3$:\\
\begin{equation}
\overline{M}_I=-1.74 + 4.5 \times ((V-I)_{\rm 0} - 1.15)\;{\rm mag}
\label{sbfrel}
\end{equation}
The sign of this slope is in agreement with the expectation that stars in blue stellar populations have higher average luminosities than stars in red populations. The value of the slope agrees reasonably well with theoretical expectations, while zero-point predictions from SBF models scatter significantly around the value Tonry obtained (see also Sect.~\ref{sbfmodels} and the discussion in Tonry et al.~\cite{Tonry01}). A limitation of the Tonry calibration is that for observational reasons it is restricted to rather luminous and large galaxies. It does not include any dwarf galaxies and does therefore exclude the blue colour range. The advantage of a calibration for bluer colours is the high number of dwarf galaxies as compared to the sparser red giant galaxies (e.g. Karick et al.~\cite{Karick03}, Peng et al.~\cite{Peng05}, Hilker et al.~\cite{Hilker03}), increasing a lot the number of accessible independent distance estimators. This gain of measurement points is partially compensated by the faint surface brightness and smaller intrinsic size of dEs, which lowers the measurement accuracy (see Sect.~\ref{sbferrors}).\\
In Mieske \& Hilker~(\cite{Mieske03b}) and Mieske et al.~(\cite{Mieske05}) we derived $I$-band SBF distances to the Centaurus and Hydra clusters and discussed how the Great Attractor hypothesis (e.g. Lynden-Bell et al.~\cite{Lynden88}, Tonry et al.~\cite{Tonry00}) fits into the derived peculiar velocities. These measurements were mainly based on distances to dEs with colours $(V-I)_0<1.10$ mag, i.e. close to or bluer than the blue limit of the Tonry calibration. A 50\% shallower calibration relation than the Tonry one was assumed for the bluest dEs, inspired by the scatter of theoretical models in that colour range (see also Sect.~\ref{sbfmodels}). The main result of our studies was that the Hydra cluster is somewhat closer to us or at most at the same distance as Centaurus. The correspondingly low Centaurus  and high Hydra peculiar velocity was consistent with a Great Attractor in close projection to and slightly behind Hydra. An empirical or semi-empirical SBF calibration in the blue regime would be very helpful to improve the accuracy of the derived distances, especially the relative one between Hydra and Centaurus.\\
Jerjen et al. (\cite{Jerjen98}, \cite{Jerjen00}, \cite{Jerjen01}) derived a semi-empirical SBF calibration between $\overline{M}_R$ and $(B-R)$ in the range $1.0<(B-R)<1.3$ mag by targeting nearby dwarf elliptical galaxies with known distances. These authors based their calibration on the theoretical predictions of Worthey~(\cite{Worthe94}) models and used their observational data to adjust the zero-point of those models. In Jerjen et al.~(\cite{Jerjen01}) they found that the models predicted too faint SBF amplitudes by 0.13 mag. Furthermore, they propose two separate branches in the $\overline{M}_R$-$(B-R)_{\rm 0}$ plane at blue colours. At a given colour, $\overline{M}_R$ becomes brighter with younger and more metal-rich stellar population. This means that the age-metallicity degeneracy of $\overline{M}_R$ is partially broken at blue colours.\\
In the course of the ACS Virgo Cluster survey, Mei et al.~(\cite{Mei05}) found a significant flattening of the SBF-colour relation in $\overline{M}_z$ vs. $(g-z)$ for $1.0<(g-z)<1.3$. They obtained a significant slope in that colour range (4.5$\sigma$), but did not find evidence for a bimodality of $\overline{M}_z$ values.\\
In the infrared, Jensen et al. (\cite{Jensen03}) presented an HST based SBF calibration between $\overline{M}_{}\rm F160W$ and $(V-I)_0$ in the range $1.05<(V-I)_0<1.24$ mag, which was based only on giant galaxies and yielded a comparable slope to that in equation~(\ref{sbfrel}). For ground based infrared bands, Jensen et al.~(\cite{Jensen98}) and Liu et al.~(\cite{Liu02}) presented calibrations between $K$ and $(V-I)_0$, also based purely on giant galaxies. The small sample of Jensen et al. in the range $1.15<(V-I)_0<1.27$ mag was consistent with a zero slope between $(V-I)_0$ and $\overline{M}_K$. Liu et al. obtained a significant slope of 3.6 $\pm$ 0.8 between both entities in the range $1.05<(V-I)_0<1.25$ mag.
\subsection{Aim of this paper}
The aim of this paper is to extend the $I$-band SBF calibration by Tonry et al. to blue colours and at the same time to test competing SBF models in that colour range. To this end we present SBF data for 25 Fornax cluster dwarf elliptical galaxies (dEs) in the range $-16.5<M_V<-11.2$ mag, $0.8<(V-I)_0<1.10$ mag.\\
The paper is structured as follows: in Sect.~\ref{sbfmodels} we compare several SBF model predictions and correspondingly define four calibration test cases. Sect.~\ref{sbfdata} presents the technical details of the SBF measurements. In Sect.~\ref{testsbf}, the four calibration test cases are compared to the real SBF data and a purely empirical calibration is derived. Sect.~\ref{discussion} contains the discussion. We finish this paper with the conclusions in Sect.~\ref{conclusions}.
\section{SBF models}
\label{sbfmodels}
\begin{figure*}[]
  \begin{center}
  \epsfig{figure=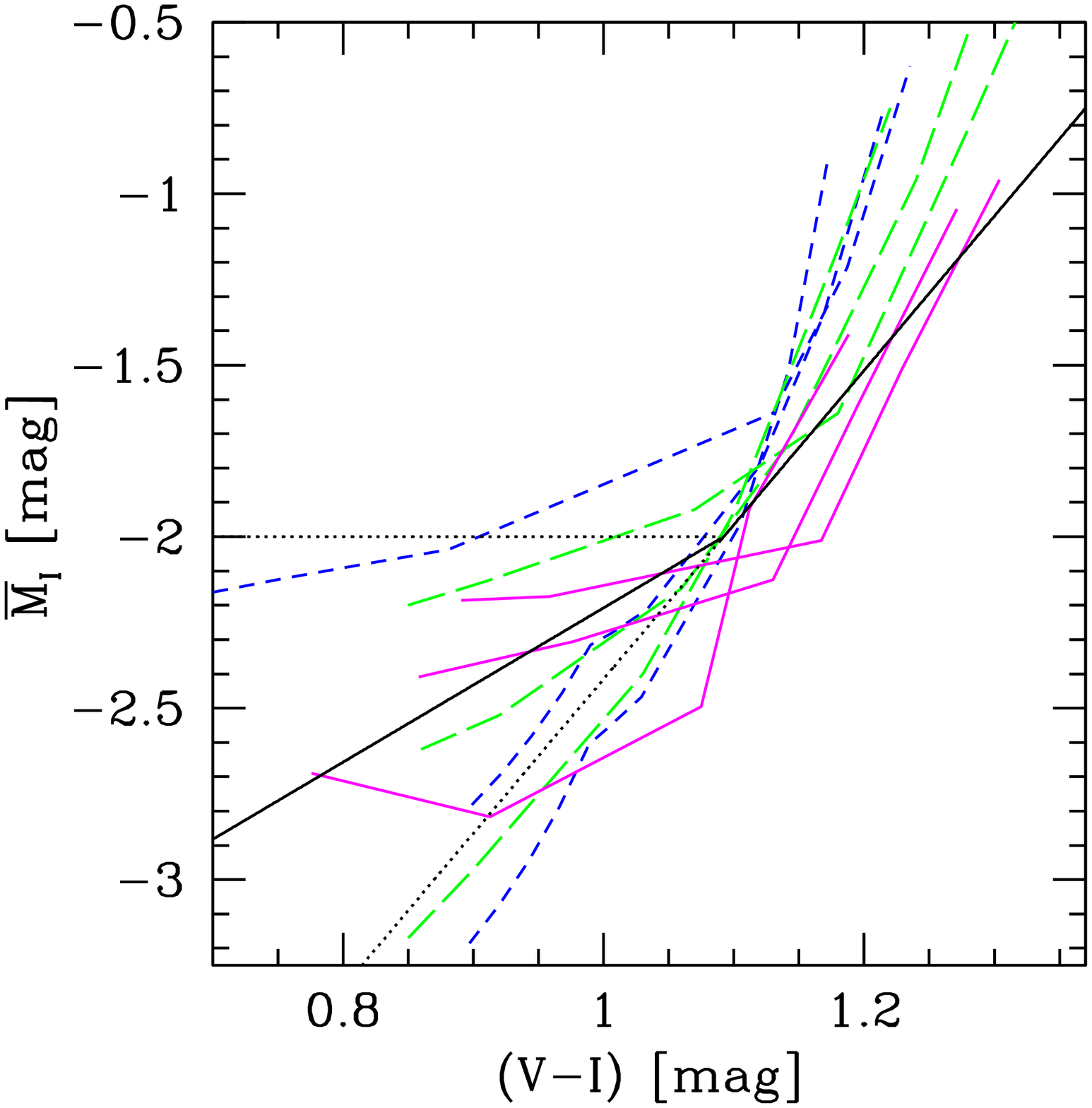,width=8.6cm}
  \epsfig{figure=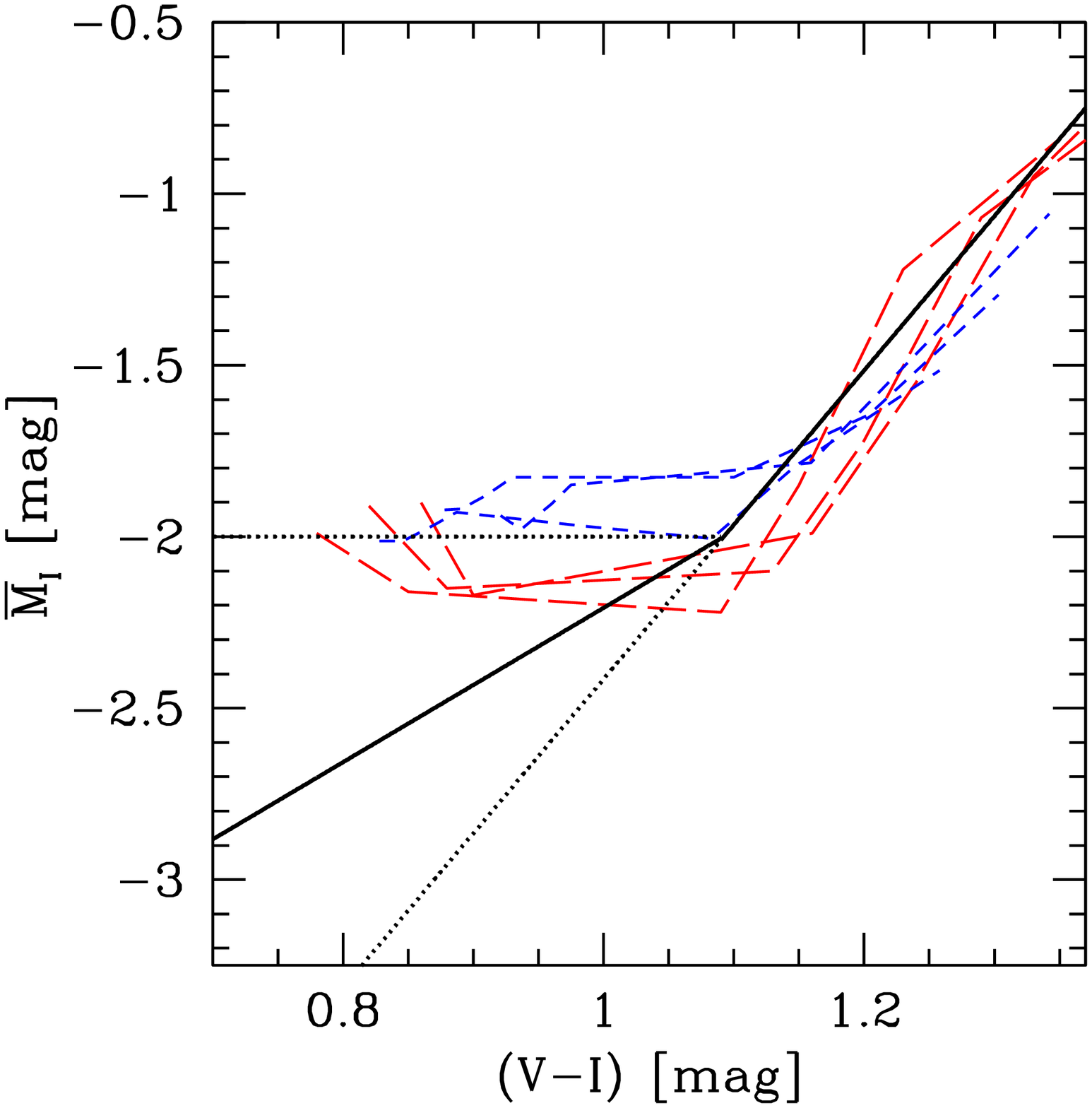,width=8.6cm}
     \caption{Theoretical ``isochrones'' in the $\overline{M}_I$-$(V-I)_{\rm 0}$ diagram from 5 different
sources. For the sake of clarity, those were separated into two different plots. {\it Left:} Blue short dashed lines from Worthey (\cite{Worthe94}) using Padova isochrones from Bertelli et al.~\cite{Bertel94} (http://astro.wsu.edu/worthey/dial/dial\_a\_pad.html). Plotted are (from top to bottom in the blue range): 17, 12 and 8 Gyrs. Metallicities range from -1.7 to 0 dex from blue to red colours. Green long dashed lines from Blakeslee et al.~(\cite{Blakes01}) for ages of 7.9, 12.6 and 17.8 Gyrs and metallicities from -1.7 to +0.2 dex. Magenta solid lines from Cantiello et al.~(\cite{Cantie03}) for ages of 5, 11 and 15 Gyrs and metallicities from -2.3 dex to 0.3 dex. 
{\it Right:} Blue short dashed lines from Worthey~(\cite{Worthe94}, http://astro.wsu.edu/worthey/dial/dial\_a\_model.html). Plotted are 17, 12 and 8 Gyrs. Metallicities range from -1.7 to 0 dex from blue to red colours. Red long dashed lines from Liu et al.~(\cite{Liu00}) for ages 8, 12 and 17 Gyrs and metallicities between -2.3 and 0.4 dex. In both panels, the solid line for $(V-I)_{\rm 0}\ge1.09$ indicates the empirical calibration ~(\ref{sbfrel}) by Tonry et al.~(\cite{Tonry97}), the solid line for $(V-I)_{\rm 0}\le1.09$ indicates equation~(\ref{sbfrelmod}), whose slope is half that of equation ~(\ref{sbfrel}). The dashed lines
indicate a constant $\overline{M}_I=-2.00$ mag and the continuation of 
equation~(\ref{sbfrel}) for $(V-I)_{\rm 0}\le1.09$. See Sect.~\ref{testsbf} for more details. }
\label{twobranch}
\end{center}
\end{figure*}

SBF measurements of low surface brightness dEs have rather high error ranges, both in colour and in SBF magnitude, see Sect.~\ref{meassbffdf} and Mieske et al.~(\cite{Mieske03a}). Because of this we do not only aim at an empirical SBF calibration, but will also test existing SBF model predictions with our measured SBF amplitudes, i.e. a semi-empirical approach.\\
In Fig.~\ref{twobranch}, theoretical values for $\overline{M}_I$ are plotted
vs. $(V-I)_{\rm 0}$ for a set of old to intermediate age stellar populations with a wide 
range of metallicities, taken from different literature sources: Worthey~(\cite{Worthe94}), who uses his own stellar population models based on VandenBerg~(\cite{Vanden85}) and Yale~(Green et al.~\cite{Green87}) isochrones; Liu et al.~(\cite{Liu00}), who use the updated models by Bruzual \& Charlot~(\cite{Bruzua93}) plus the Bertelli et al.~(\cite{Bertel94}) Padova isochrones; Blakeslee et al.~(\cite{Blakes01}) who use the stellar population models of Vazdekis et al.~(\cite{Vazdek96}) and the updated Girardi et al.~(\cite{Girard00}) Padova isochrones; and Cantiello et al.~(\cite{Cantie03}), who use the stellar population synthesis code by Brocato et al.~(\cite{Brocat99},~\cite{Brocat00}) and the Teramo-Pisa-Rome isochrones (e.g. Castellani et al.~\cite{Castel91}, Castellani et al.~\cite{Castel92} and Bono et al.~\cite{Bono97a},~\cite{Bono97b}). Equation~(\ref{sbfrel}) 
is also shown in Fig.~\ref{twobranch}. The Worthey model predictions are split up into the original predictions from Worthey~(\cite{Worthe94})\footnote{http://astro.wsu.edu/worthey/dial/dial\_a\_model.html} and those that use the alternate  stellar evolutionary isochrone library from Bertelli et al.~(\cite{Bertel94})\footnote{http://astro.wsu.edu/worthey/dial/dial\_a\_pad.html}. For the sake of clarity, we separate the model predictions into two different plots.\\
For red colours, all models trace equation~(\ref{sbfrel}) reasonably well, with the original Worthey-models and those of Blakeslee et al.
being more
deviant from the empirical calibration than the other models. In the blue range, there is a notable discrepancy between the model predictions: the original Worthey~(\cite{Worthe94}) models and those of Liu et al.~(\cite{Liu00}) predict a colour independent $\overline{M}_I\simeq -2.0$ mag for $(V-I)_{\rm 0}<1.10$ mag. In contrast, Worthey models using the Bertelli et al.~(\cite{Bertel94}) Padova isochrones, and also the models of Blakeslee et al.~(\cite{Blakes01}) and Cantiello et al.~(\cite{Cantie03}) show a large spread in $\overline{M}_I$ caused by age differences. Those models predict a continuation of equation~(\ref{sbfrel}) for intermediate ages and a flattening of the relation for very old ages. At $(V-I)_{\rm 0}\simeq 0.90$ this leads to an uncertainty 
of up to 0.5 mag in relating $\overline{M}_I$ to $(V-I)_{\rm 0}$. \\
Based on the different theoretical model predictions in Fig.~\ref{twobranch}, we define four ``test cases'' for possible calibrations:\\
{\bf Case A:} This is the ``original'' Tonry calibration, equation~(\ref{sbfrel}), applied to all galaxies.\\
{\bf Case B:} This is equation~(\ref{sbfrel}) applied for $(V-I)_0>1.09$, and the following equation~(\ref{sbfrelmod}) -- which has half the slope of equation~(\ref{sbfrel}) -- applied for $(V-I)_0<1.09$:
\begin{equation}
\overline{M}_I=-2.0 + 2.25 \times ((V-I)_{\rm 0} - 1.09)\;{\rm mag}
\label{sbfrelmod}
\end{equation}
This case represents the theoretical SBF models plotted in the left panel of Fig.~\ref{twobranch}, which viewed as a whole suggest a non-zero but shallower slope than that of the empirical calibration at redder colours. See case C) below for the reason of choosing $(V-I)=1.09$ as colour limit.\\
{\bf Case C:} This is equation~(\ref{sbfrel}) applied for $(V-I)_0>1.09$ and a constant of $\overline{M}_I=-2.00$ mag applied for $(V-I)_0<1.09$. This represents the theoretical SBF models plotted in the right panel of Fig.~\ref{twobranch}. The specific limit of $(V-I)_0=1.09$ is chosen because at this colour, the colour independent $\overline{M}_I\simeq -2.0$ mag of the models in the right panel of Fig.~\ref{twobranch}  matches the value of $\overline{M}_I$ from equation~(\ref{sbfrel}). This is also the reason for adopting the same limit in case B), since the latter case represents the ``hybrid'' solution between case A) and case C).\\
{\bf Case D:} A flat and a steep calibration relation are valid in the same colour regime. Depending on the strength of the SBF signal, one of these is applied to a given galaxy. This case represents the ``two branch'' hypothesis put forward in the investigations by Jerjen et al. The steep relation is equation~(\ref{sbfrel}). It is applied to galaxies which have $(V-I)_0>1.09$ {\it or} whose distance in case B) is smaller than the mean distance. The flat relation is simply a constant $\overline{M}_I=-2.00$ mag, like in case C). This is applied to all galaxies with $(V-I)_0<1.09$ whose distance in case B) is larger than the mean distance. In other words, galaxies with $\overline{m}_I$ above the line for case B) in Fig.~\ref{twobranch} are assigned $\overline{M}_I=-2.00$ mag, while those below are assigned relation~(\ref{sbfrel}). The mean distance obtained with this case is of course very close to that in case B). \\
Those four cases will be tested against the real SBF measurements in Sect.~\ref{testsbf}.
\section{The SBF data}
\label{sbfdata}
The observations for this paper were taken in the nights 25th of October 2003 and 5-7 of December 2004 at Las Campanas Observatory (LCO), Chile. The instrument used was the Inamori
Magellan Areal Camera and Spectrograph ``IMACS'' in imaging mode with the
``short'' f/2 camera, mounted at the 6.5m Baade telescope. 
The double-asphere, glass-and-oil-lens f/2 camera
produces an image 
of 27.4$'$ field diameter at 0.20 arcsec per pixel. The field is vignetted
in the corners (by the tertiary mirror and its mounting assembly), changing from
0\% flux loss at R = 12$'$ to ~10\% at R = 15$'$. The IMACS detector is an 8192$\times$8192
CCD mosaic camera which uses 8 thinned 2k$\times$4k detectors. 
Gaps of about 50 pixels separate the chips.\\
In total, 7 fields in the central Fornax cluster were observed in the two bands $V$ and $I$. The total integration times in $V$ were between 1200 and 1800 seconds except for the central field around NGC 1399, for which it was 900 seconds. 
The total $I$-band integration time was between 3800 and 5160 seconds, except for the central NGC 1399 field, which had 1800 seconds integration.\\
Standard star images of the Landolt fields SA92, SA95, SA98 and SA104 -- the latter one only in the second run -- were taken at different airmass values in one short (2 seconds) and one long exposure (10 seconds). This resulted in about 110 data points for the first run and about 140 for the second one, each time corresponding to about 40 standard stars. Comparison of calibrated colours and magnitudes for objects imaged in both runs yielded absolute zero point uncertainties of the order 0.02 mag.
\subsection{Image reduction before SBF measurement}
The image reduction steps before the SBF measurement were the following: first, a master-bias was created for each chip and was subtracted from the domeflat exposures. Then for each chip the bias corrected dome flats were combined to master-domeflats. Having the master-biases and master-domeflats prepared for each chip, we used the COSMOS package\footnote{Carnegie Observatories System for
 MultiObject Spectroscopy, http://llama.lco.cl/aoemler/COSMOS.html} to do bias-subtraction, trimming and flat-field correction of the raw science frames. The reduced single science frames were registered with integer pixel shifts to avoid distortion of the SBF power spectrum and combined using a min-max rejection algorithm. The seeing FWHM of the combined images  generally ranged between 0.6 and 0.9$''$, except at the edges of the field of view: especially for the first run in 2003, when the atmospheric dispersion corrector had not been installed yet, distortions close to the image border caused seeing degradations to about 2$''$. There were two Fornax dEs for which we could not measure SBF because they were located in these regions of high distortion.\\
The same image reduction procedure as for the single science frames was performed for the standard star images. Their instrumental magnitudes were measured with the IRAF package APPHOT in apertures equal to those used by Landolt~(\cite{Landolxx}). Then, a single photometric solution was determined for the entire 8k$\times$8k image. Chip-to-chip variations around the mean photometric zero point were smaller than 0.02 mag. To correct for galactic reddening and 
absorption, we used the values from Schlegel et al. (\cite{Schleg98}), who give $A_I=0.025$ and 
$E(V-I)=0.018$ for the coordinates of the Fornax cluster.
\subsection{SBF measurement}
\label{meassbffdf}
\begin{figure*}[]
   \begin{center}
   \epsfig{figure=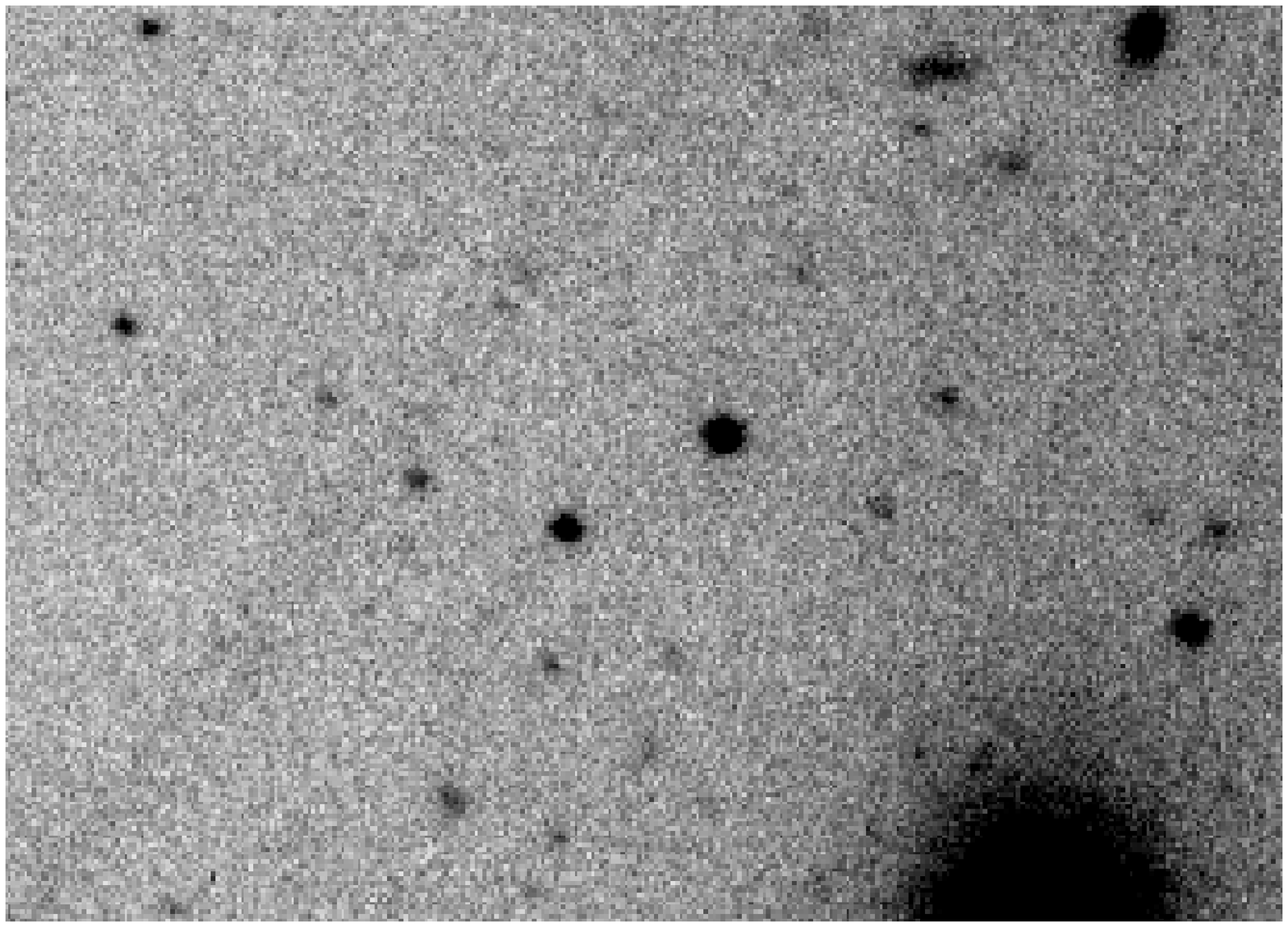,width=8.6cm}
   \epsfig{figure=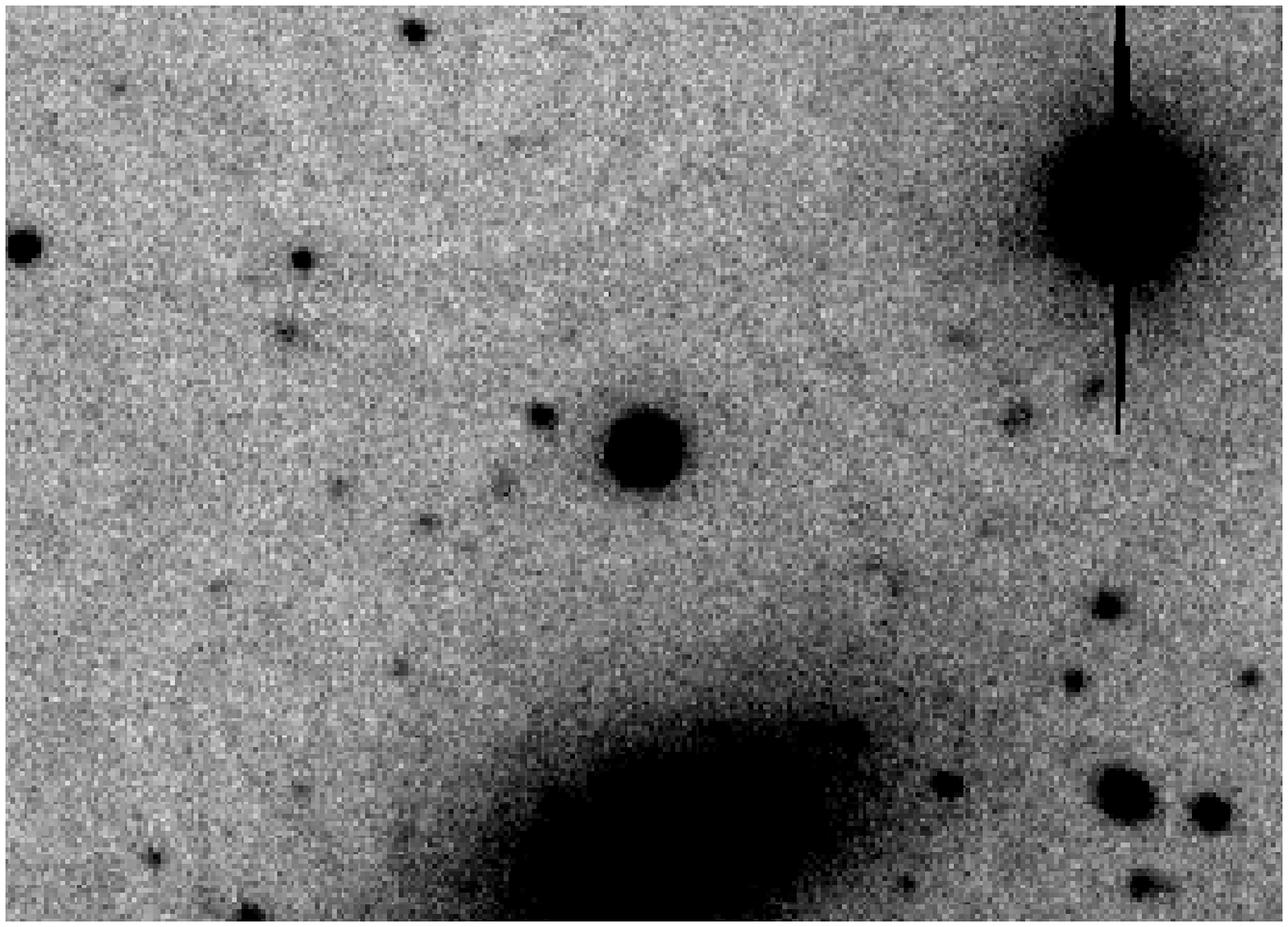,width=8.6cm}
      \caption{Two image excerpts from the same chip and exposure close to two investigated dEs, illustrating the variable degree of fringing. In the left thumbnail, FCC 194 is at the bottom right. In the right thumbnail, FCC 196 is at the bottom. Both images have the same intensity cuts. Note the fringing patterns in the right thumbnail, which are not present in the left one. }
         \label{fringing}
\end{center}
   \end{figure*}

\begin{figure}[]
   \begin{center}
   \epsfig{figure=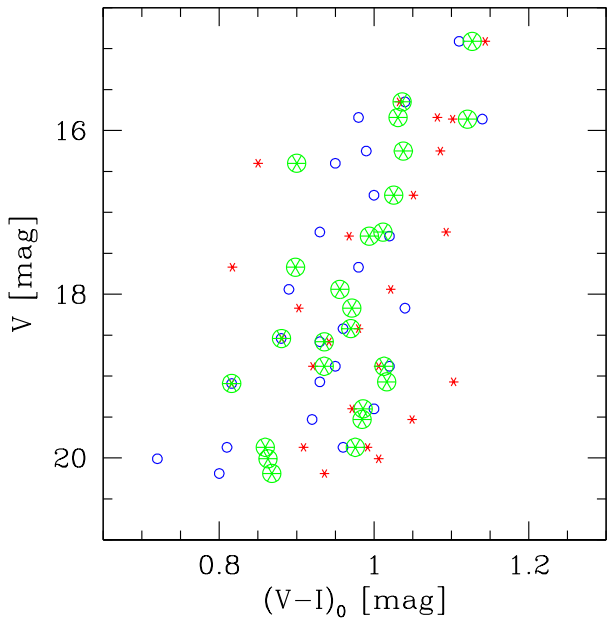,width=8.6cm}
      \caption{Colour magnitude diagram of the galaxies entering our SBF analysis, i.e. those ones with an SBF S/N $>3$, see also Fig.~\ref{biastest}. Small blue circles indicate colours derived from the present IMACS data set. Small red asterisks indicate colours from the WFCCD data set of Hilker et al.~(\cite{Hilker03}). Large encircled asterisks is the mean of both values. The $V$ magnitude is from the WFCCD data set for all galaxies.}
         \label{CMD}
\end{center}
   \end{figure}

\begin{figure}[]
   \begin{center}
   \epsfig{figure=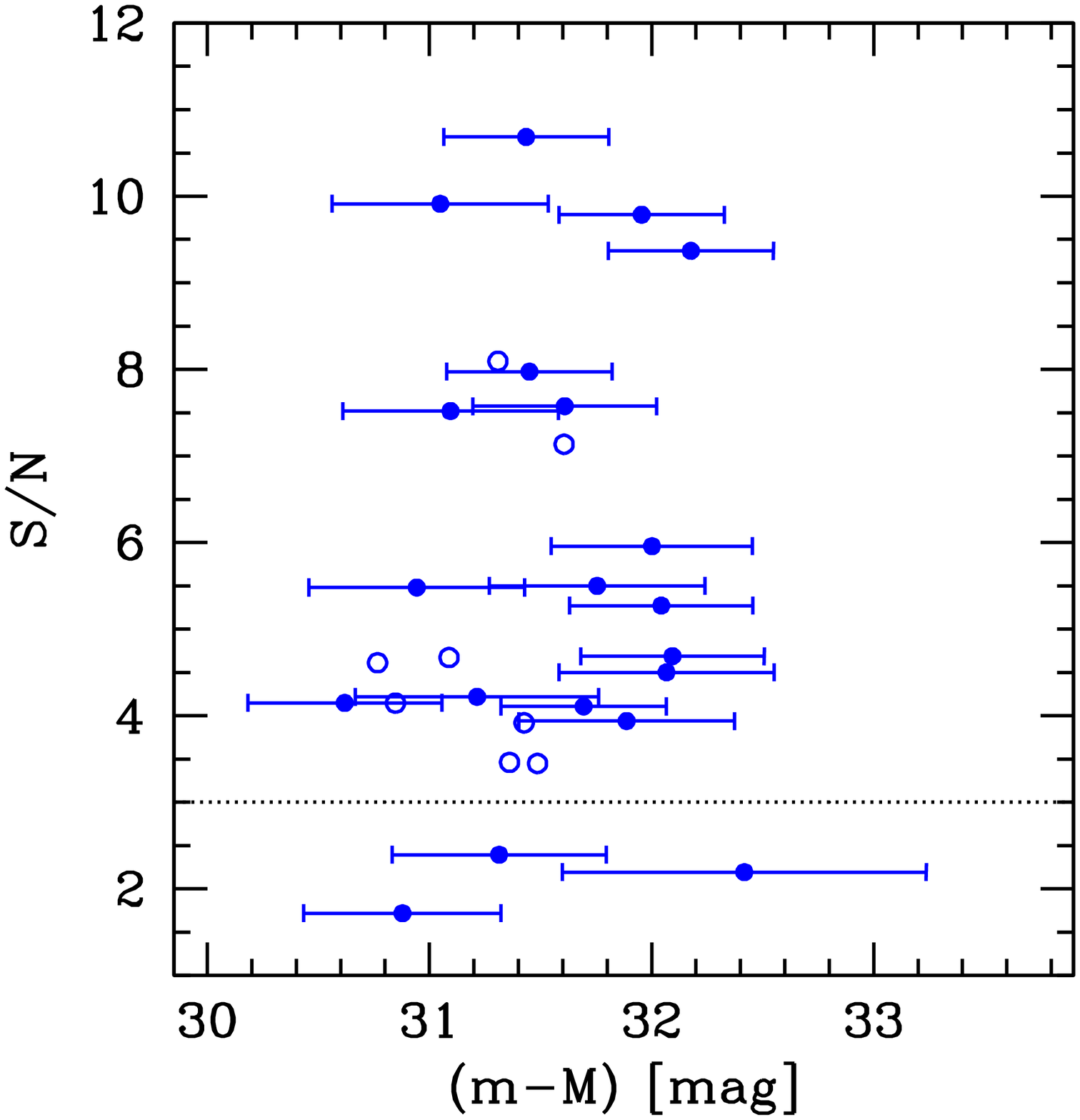,width=4.4cm}\hspace{-0.1cm}
\epsfig{figure=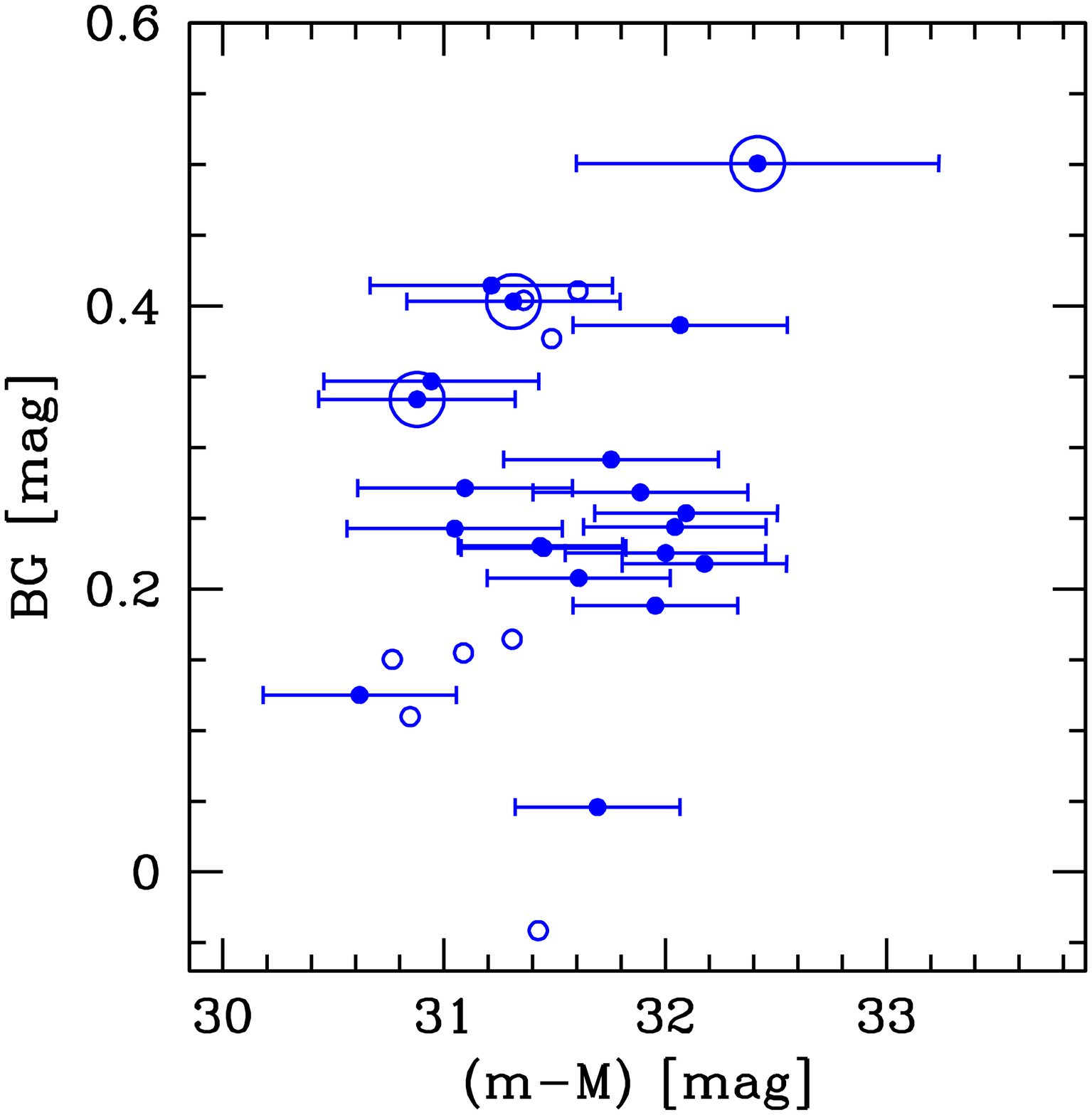,width=4.4cm}
        \caption{Distance modulus for case A) (see text) plotted vs. S/N of the SBF measurement (left) and the amount of background fluctuations $BG$ contributing to the measured SBF signal (right). The three sources marked with large circles in the right panel are those below $S/N=3$ in the left panel. Filled circles are for galaxies with seeing below 0.85$''$, open circles for seeing above 0.85$''$, see Fig.~\ref{seeing} and text.}
         \label{biastest}
\end{center}
   \end{figure}

The SBF measurement procedure was similar to that already outlined in Mieske et al.~(\cite{Mieske05}) based on VLT FORS1 photometry. Some nuances have been modified: the fluctuations from the sky background were now derived in the same image close to the investigated dE instead of in a comparison field. Those sky fluctuations consisted of undetected background galaxies and fringing. The seeing was not constant over one chip, generally increasing towards the edges of the field of view. This caused a correspondingly varying amount of fluctuation from undetected background galaxies along the chip. Also the fringing varied in some cases significantly over one chip, see Fig.~\ref{fringing}. These two dependences on position required a local determination of sky fluctuations. Note that for the VLT FORS photometry there was no significant fringing nor strong variations of seeing among the different exposures.\\
In turn we describe the relevant measurement steps for determining the SBF magnitude:\\
1.~Model mean galaxy light with ELLIPSE using a sigma clipping algorithm to disregard
contaminating sources, subtract the model.\\
2.~Detect and subtract remaining contaminating objects from original image.\\
3.~Model mean galaxy light on the cleaned image.\\
4.~Subtract model of original image.\\
5.~Divide resulting image by square root of the model, cut out circular portion
with radius typically 20-25 pixel ($4-5''$), corresponding to about 4-8 seeing disk diameters.\\
6.~Mask out contaminating sources like foreground stars and background galaxies.\\
7.~Calculate the power spectrum (PS) of the cleaned image.\\
8.~Calculate the PS of the sky background in three different blank image sections close to the investigated galaxy. This sky background PS contains both fluctuation contributions from undetected background galaxies as fluctuations caused by fringing, see explanation at beginning of this subsection. Normalise these fluctuations by dividing these blank images by the mean galaxy intensity in the region where SBF are measured. Subtract this normalised background PS $BG$ from the PS of the SBF image.\\
9.~Obtain the azimuthal average of the resulting PS.\\
10.~Fit function of the form
\begin{equation}
\label{azimut2}
P(k)=E(k)\times P_{\rm 0}+P_{\rm 1}
\end{equation}
to the result. Here, $E(k)=PSF(k) \otimes W(k)$ with $W(k)$ the PS of the mask used to excise contaminating sources and restrict the measurement area. 
$PSF(k)$ is the 
undistorted power spectrum of the PSF normalised to unity at k=0. $PSF(k)$ is derived from one to three well exposed star(s) in the same image close to the respective galaxy. When more than one star was used, the average PSF-PS was adopted as $PSF(k)$. The variations in PSF FWHM between stars chosen for one galaxy were small, around 5\%. Using $E(k)$ instead of 
$PSF(k)$ implicitly corrects for the power spectrum damping at low wavenumbers that is caused by
the convolution of the undistorted $PSF(k)$ with the power spectrum of the mask (see Mieske et al.~\cite{Mieske05}). The amount of power spectrum damping at wavenumber zero was on average 0.11 mag with a galaxy-to-galaxy dispersion of 0.04 mag. The signal-to-noise ratio of the SBF measurement is then defined as $S/N=\frac{P_0}{P_1}$. Since the background fluctuations have been subtracted from the power spectrum before the fitting, this definition corresponds to the ratio of pure SBF signal over sky noise. For the SBF analysis, only galaxies with $S/N>3$ were considered (see for example Figs.\ref{CMD}~and~\ref{biastest}).\\
11.~Obtain the desired observable $\overline{m}_I$ from 
\begin{equation}
\label{mbarI2}
\overline{m}_I=-2.5*log(P_{\rm 0}) + ZP_I - A_I - \Delta k + \Delta_{\rm GC}
\end{equation}
$A_I$ is
the foreground absorption,
$\Delta k=z\times 7$ is the k-correction for SBF in the $I$-band (Tonry et al. \cite{Tonry97}). For calculating $\Delta k$ we assumed $cz=1500$ km/s for all galaxies. $\Delta_{\rm GC}$ is the fluctuation contribution from undetected globular clusters (GCs). To estimate $\Delta_{\rm GC}$, we determined the completeness magnitude for detecting unresolved point sources for each galaxy separately according to the seeing via artificial star experiments. The 50\% completeness magnitudes ranged between $I$=22.5 and 24.2 mag. For calculating $\Delta_{\rm GC}$, we assumed equation~(15) from Blakeslee \&
Tonry~(\cite{Blakes95}), a turn-over magnitude (TOM) of $M_I=-8.46$ mag and Gaussian width $\sigma=1.2$ mag for the GCLF (Kundu \& Whitmore \cite{Kundu01}), a Fornax distance modulus of 31.4 mag, and a specific frequency of its GC system $S_N:=\frac{N_{GC}}{10^{-0.4*(M_V+15)}}=5$ (Miller et al.~\cite{Miller98}). The calculated values for $\Delta_{\rm GC}$ are listed in Table~\ref{sbfresults}, ranging between 0.01 mag and 0.42 mag with a mean of 0.06 mag and median of 0.03 mag.\\
The colour $(V-I)_{\rm 0}$ of each galaxy was adopted as the mean of two values: first, the $(V-I)_{\rm 0}$ estimates from the IMACS data in the region where SBF were measured; second, the colours derived by us for the same galaxies in Hilker et al.~(\cite{Hilker03}). These latter data were based on WFCCD photometry with the 2.5m du Pont telescope, also at Las Campanas. Those colours had been determined in circular apertures of 4$''$ radius, which is in close agreement to the size of the region where SBF were measured. For those galaxies entering our SBF data set, the systematic difference between the IMACS and WFCCD colours is 0.017 $\pm$ 0.021 mag, with a scatter of 0.10 mag. This shows that our photometric colour calibration is accurate to better than 0.02 mag. Fig.~\ref{CMD} shows the CMD of the investigated galaxies for the various colour sets. The rms dispersion in colour of a linear fit is 0.083 mag for the IMACS data, 0.068 for the WFCCD data and 0.058 mag for the mean, independent of magnitude. We adopt this very dispersion of 0.058 mag as colour uncertainty. This is to be seen as an upper limit to the real colour error because the rms dispersion certainly includes some ``cosmic'' scatter.\\
With $(V-I)_{\rm 0}$ at hand, $\overline{M}_I$ was derived using equation~(\ref{sbfrel}), which is calibration case A). Together with the measured $\overline{m}_I$, this yielded the ``null hypothesis'' distance modulus $(m-M)_A$, see Table~\ref{sbfresults}. Fig.~\ref{thumbnails} shows thumbnail images illustrating the SBF measurement procedure for two out of the 28 dEs with clear SBF signal. 25 of the 28 investigated galaxies had $S/N>3$ and entered our subsequent analysis. Their absolute brightnesses are in the range $-16.5<M_V<-11.2$ mag, their central surface brightnesses cover the regime $20.3<\mu_V<24.6$ mag/arcsec$^2$. They occupy a colour range $0.8<(V-I)_0<1.10$ mag, ideally suited for extending the Tonry colour range to the blue. In Fig.~\ref{biastest} we show that the null hypothesis SBF distance modulus does not correlate with the S/N of the SBF measurement nor with the amount of background fluctuations $BG$. 
%This makes us confident that there are no strong observational biases in our SBF measurements.
\\ Note the large range of $BG$ values in Fig.~\ref{biastest}. This has three reasons: The first and main reason is related to the fact that $BG$ is determined by fitting equation~(\ref{azimut2}) to the normalised sky background PS. For this fitting we have to exclude the same low wavenumber regime that is excluded for fitting the galaxy SBF amplitude $P_0$. Since $BG$ is only a fraction of $P_0$, this exclusion of low wavenumbers has a stronger effect on the fitting precision for $BG$ than for fitting $P_0$. Indeed, the  negative  value for $BG$ that is measured for FCC 215 (see Table~\ref{sbfresults} and Fig.~\ref{biastest}) is a consequence of this low wavenumber exclusion. The background PS in the remaining wavenumbers is predominantly noise dominated such that a formally negative value for $BG$ is fitted. As a consequence of that, the SBF data with poor seeing (FWHM$>$0.85$''$) show a scatter in $BG$ of 0.16 mag which is almost twice as large as the scatter of 0.09 mag for the data with good seeing  (FWHM$<$0.85$''$), see Fig.~\ref{biastest}. The second reason for the variation of $BG$ is that the galaxies investigated span a considerable range in surface brightness ($>$ 4 mag). This leads to varying relative amounts of $BG$, given that the SBF signal in lower surface brightness regions is more strongly affected by sky background fluctuations. The third reason is the somewhat varying fringing amplitude, both between different chips and different nights, see for example Fig.~\ref{fringing}.\\
We finally note that the sky background {\it correction} of the galaxy SBF data is done by {\it directly subtracting} the sky background PS image from the normalised galaxy light PS image. Only the {\it quantification of the correction} is achieved by {\it fitting} a PS. To take the uncertainty arising from the power-spectrum fit and from the sky fluctuation variability into account, we adopt the scatter of $P_0$ resulting from the subtraction of three different sky background PS images as error estimate of the SBF measurement, see the following Sect.~\ref{sbferrors}.

\subsubsection{Error estimate for SBF distances}
\label{sbferrors}
The error of $(m-M)$ has two equally important contributions: uncertainty in $(V-I)_{\rm 0}$ and $P_{\rm 0}$.\\
The uncertainty in $(V-I)_0$ of 0.058 mag (see above) translates into a 0.26 mag distance uncertainty in the case of the empirically calibrated slope 4.5.\\
The error of $P_{\rm 0}$ was estimated in two ways:\\
1. based on the Monte Carlo simulations presented in Mieske et al.~(\cite{Mieske03a}). For the median seeing of our data (0.75$''$), interpolation of the simulation results for 0.5 and 1.0$''$ for Fornax dEs and 1 hour integration yields the following uncertainties: 0.37 mag in the range $-11>M_V>-13$ mag, 0.27 mag in the range $-13>M_V>-14.5$ mag, and 0.20 mag in the range $M_V<-14.5$ mag. Note that the simulation errors are calculated for images of the same pixel scale than the one used here. The fainter VLT-FORS zero points assumed in the simulations are practically compensated by the larger integration times of our Magellan-IMACS data.\\
2. from the scatter of the fitted values for $P_{\rm 0}$ when subtracting the three different sky background power spectra. This was an important double-check of the errors adopted from the simulations, because the fringing of the IMACS data represent an additional uncertainty source, apart from the photon noise and background galaxy fluctuations which had been included in the simulations.\\
The maximum of both error estimates was adopted as error in $P_{\rm 0}$. For seven out of the 25 dEs investigated, the scatter from background subtraction was larger than the Monte Carlo estimate, see Table~\ref{sbfresults}. The resulting mean error of $(m-M)$ is 0.41 mag when adopting calibration case A). There are four galaxies in the range $-16.5<M_V<-11.5$ mag which were imaged twice in the course of our survey since they were located in the overlap of adjacent fields. The differences in $(m-M)$ between those double measurements ranged between 0.10 and 0.35 mag. This is encouragingly low, indicating that our uncertainty estimates are on the conservative side.
\begin{table*}
\begin{tabular}{r|rrrrrrrrrrr}
Name & $(V-I)_0$& $V$& $\overline{m}_I$& S/N & $BG$ & {\bf $\Delta_{\rm GC}$} & seeing [$''$] &$(m-M)_A$ &$(m-M)_B$ &$(m-M)_C$ &$(m-M)_D$\\\hline\hline
FCC 222 & 1.127 & 14.91 & 29.61 $\pm$ 0.20&   8.0 & 0.23 &  0.08 & 0.72 & 31.45 $\pm$ 0.33 & 31.45 $\pm$ 0.33 & 31.45 $\pm$ 0.33 & 31.45 $\pm$ 0.33 \\
FCC 211 & 1.036 & 15.65 & 29.44 $\pm$ 0.20& 4.1 & 0.05 &  0.07 & 0.65 & 31.69 $\pm$ 0.33 & 31.57 $\pm$ 0.24 & 31.45 $\pm$  0.20 & 31.45 $\pm$  0.20 \\
FCC 223 & 1.031 & 15.84 & 28.57 $\pm$ 0.20& 4.1 & 0.11 &  0.29 & 0.86 & 30.85 $\pm$ 0.33 & 30.71 $\pm$ 0.24 & 30.58 $\pm$  0.20 & 30.85 $\pm$ 0.33 \\
FCC 188 & 1.121 & 15.86 & 29.56 $\pm$ 0.20&  11.0 & 0.23 &  0.02 & 0.71 & 31.44 $\pm$ 0.33 & 31.44 $\pm$ 0.33 & 31.44 $\pm$ 0.33 & 31.44 $\pm$ 0.33 \\
FCC 241 & 1.038 & 16.25 & 28.52 $\pm$ 0.20& 4.6 & 0.15 &  0.04 &    1.00 & 30.77 $\pm$ 0.33 & 30.65 $\pm$ 0.24 & 30.53 $\pm$  0.20 & 30.77 $\pm$ 0.33 \\
FCC 274 & 0.900 & 16.40 & 29.31 $\pm$ 0.20& 9.4 & 0.22 &  0.03 & 0.65 & 32.18 $\pm$ 0.33 & 31.75 $\pm$ 0.24 & 31.32 $\pm$  0.20 & 31.32 $\pm$  0.20 \\
FCC 156 & 1.026 & 16.79 & 29.66 $\pm$ 0.20& 9.8 & 0.19 &  0.06 & 0.66 & 31.96 $\pm$ 0.33 & 31.81 $\pm$ 0.24 & 31.67 $\pm$  0.20 & 31.67 $\pm$  0.20 \\
FCC 160 & 1.012 & 17.24 & 29.24 $\pm$ 0.31& 7.1 & 0.41 &  0.42 &  1.20 & 31.61 $\pm$ 0.41 & 31.43 $\pm$ 0.34 & 31.25 $\pm$ 0.31 & 31.25 $\pm$ 0.31 \\
FCC 196 & 0.994 & 17.29 & 29.65 $\pm$ 0.27& 4.7 & 0.25 & 0.03 & 0.67 & 32.09 $\pm$ 0.37 & 31.88 $\pm$  0.30 & 31.66 $\pm$ 0.27 & 31.66 $\pm$ 0.27 \\
FCC 287 & 0.898 & 17.67 & 29.17 $\pm$ 0.27& 5.3 & 0.24 &  0.07 & 0.79 & 32.04 $\pm$ 0.37 & 31.61 $\pm$  0.30 & 31.18 $\pm$ 0.27 & 31.18 $\pm$ 0.27 \\
FCC 218 & 0.956 & 17.94 &    28.00 $\pm$ 0.30& 4.1 & 0.13 &  0.04 &  0.80 & 30.62 $\pm$ 0.40 & 30.32 $\pm$ 0.33 & 30.01 $\pm$  0.30 & 30.62 $\pm$  0.40 \\
FCC 194 & 0.971 & 18.17 & 29.07 $\pm$ 0.27& 7.6 & 0.21 &  0.02 & 0.73 & 31.61 $\pm$ 0.37 & 31.34 $\pm$  0.30 & 31.08 $\pm$ 0.27 & 31.08 $\pm$ 0.27 \\
FCC 140 &  0.970 & 18.42 & 29.45 $\pm$ 0.33&   6.0 & 0.22 &  0.08 & 0.65 &    32.00 $\pm$ 0.42 & 31.73 $\pm$ 0.35 & 31.46 $\pm$ 0.33 & 31.46 $\pm$ 0.33 \\
FCC 229 & 0.881 & 18.54 & 29.11 $\pm$ 0.37& 4.5 & 0.39 &  0.02 & 0.67 & 32.07 $\pm$ 0.45 &  31.60 $\pm$ 0.39 & 31.12 $\pm$ 0.37 & 31.12 $\pm$ 0.37 \\
FCC 168 & 0.936 & 18.58 & 28.66 $\pm$ 0.40& 3.5 & 0.40 &  0.03 & 0.88 & 31.36 $\pm$ 0.48 & 31.01 $\pm$ 0.42 & 30.67 $\pm$  0.40 & 31.36 $\pm$ 0.48 \\
FCC 145 & 1.013 & 18.88 & 28.69 $\pm$ 0.37& 9.9 & 0.24 &  0.04 & 0.66 & 31.05 $\pm$ 0.45 & 30.88 $\pm$ 0.39 &  30.70 $\pm$ 0.37 & 31.05 $\pm$ 0.45 \\
FCC 214 & 0.936 & 18.88 & 29.18 $\pm$ 0.37& 3.9 & 0.27 &  0.03 & 0.81 & 31.89 $\pm$ 0.45 & 31.54 $\pm$ 0.39 & 31.19 $\pm$ 0.37 & 31.19 $\pm$ 0.37 \\
FCC 144 & 1.016 & 19.07 & 28.75 $\pm$ 0.37& 7.5 & 0.27 &  0.03 & 0.62 & 31.09 $\pm$ 0.45 & 30.93 $\pm$ 0.39 & 30.76 $\pm$ 0.37 & 31.09 $\pm$ 0.45 \\
FCC 215 & 0.816 & 19.09 & 28.18 $\pm$ 0.37& 3.9 & -0.04 &  0.02 & 0.88 & 31.43 $\pm$ 0.45 & 30.81 $\pm$ 0.39 & 30.19 $\pm$ 0.37 & 31.43 $\pm$ 0.45 \\
FCC 191 & 0.986 & 19.40 & 28.74 $\pm$ 0.45& 4.2 & 0.41 &  0.01 & 0.68 & 31.21 $\pm$ 0.52 & 30.98 $\pm$ 0.47 & 30.75 $\pm$ 0.45 & 31.21 $\pm$ 0.52 \\
FCC 192 & 0.985 & 19.53 & 28.46 $\pm$ 0.37& 5.5 & 0.35 &  0.01 & 0.72 & 30.94 $\pm$ 0.45 & 30.71 $\pm$ 0.39 & 30.47 $\pm$ 0.37 & 30.94 $\pm$ 0.45 \\
LSB 9-4 & 0.860 & 19.87 & 28.26 $\pm$ 0.37& 8.1 & 0.17 &  0.02 & 0.86 & 31.31 $\pm$ 0.45 & 30.79 $\pm$ 0.39 & 30.27 $\pm$ 0.37 & 31.31 $\pm$ 0.45 \\
LSB 6-4 & 0.976 & 19.87 & 28.96 $\pm$ 0.53& 3.4 & 0.38 &  0.07 & 0.98 & 31.49 $\pm$ 0.59 & 31.23 $\pm$ 0.55 & 30.97 $\pm$ 0.53 & 31.49 $\pm$ 0.59 \\
LSB 6-2 & 0.863 & 20.01 & 28.72 $\pm$ 0.37& 5.5 & 0.29 &  0.04 & 0.64 & 31.76 $\pm$ 0.45 & 31.24 $\pm$ 0.39 & 30.73 $\pm$ 0.37 & 30.73 $\pm$ 0.37 \\
LSB 10-8 & 0.868 & 20.19 & 28.07 $\pm$ 0.40& 4.7 & 0.16 &     0.01 & 0.97 & 31.09 $\pm$ 0.48 & 30.59 $\pm$ 0.42 & 30.09 $\pm$  0.40 & 31.09 $\pm$ 0.48 \\\hline
\end{tabular}
\caption{Results of SBF measurements for the 25 Fornax dEs with S/N$>3$ in the SBF measurement. The galaxies are ordered by decreasing brightness. In the first column, the reference number in the Fornax Cluster Catalog FCC (Ferguson \& Sandage~\cite{Fergus89}) is given. The last four galaxies had been too faint for this catalog. They are labelled with a field number and running index according to Hilker et al.~(\cite{Hilker03}). All columns have magnitudes as units except S/N and seeing. $(V-I)_0$ is the mean colour of the photometry from Hilker et al.~(\cite{Hilker03}) and this paper. $V$ is from Hilker et al.~(\cite{Hilker03}). The other parameters are derived in this paper. $BG$ gives the amount of background fluctuation present in the original fluctuation image, comprising both fluctuations from undetected background sources and fluctuations arising from fringing. $(m-M)_A$ to $(m-M)_D$ give the four sets of distances obtained when applying the four different calibration test cases, see Sect.~\ref{sbfmodels}.}
\label{sbfresults}
\end{table*}
\subsection{Correlation between seeing and $\overline{m}_I$}
\begin{figure}[]
\begin{center}
\epsfig{figure=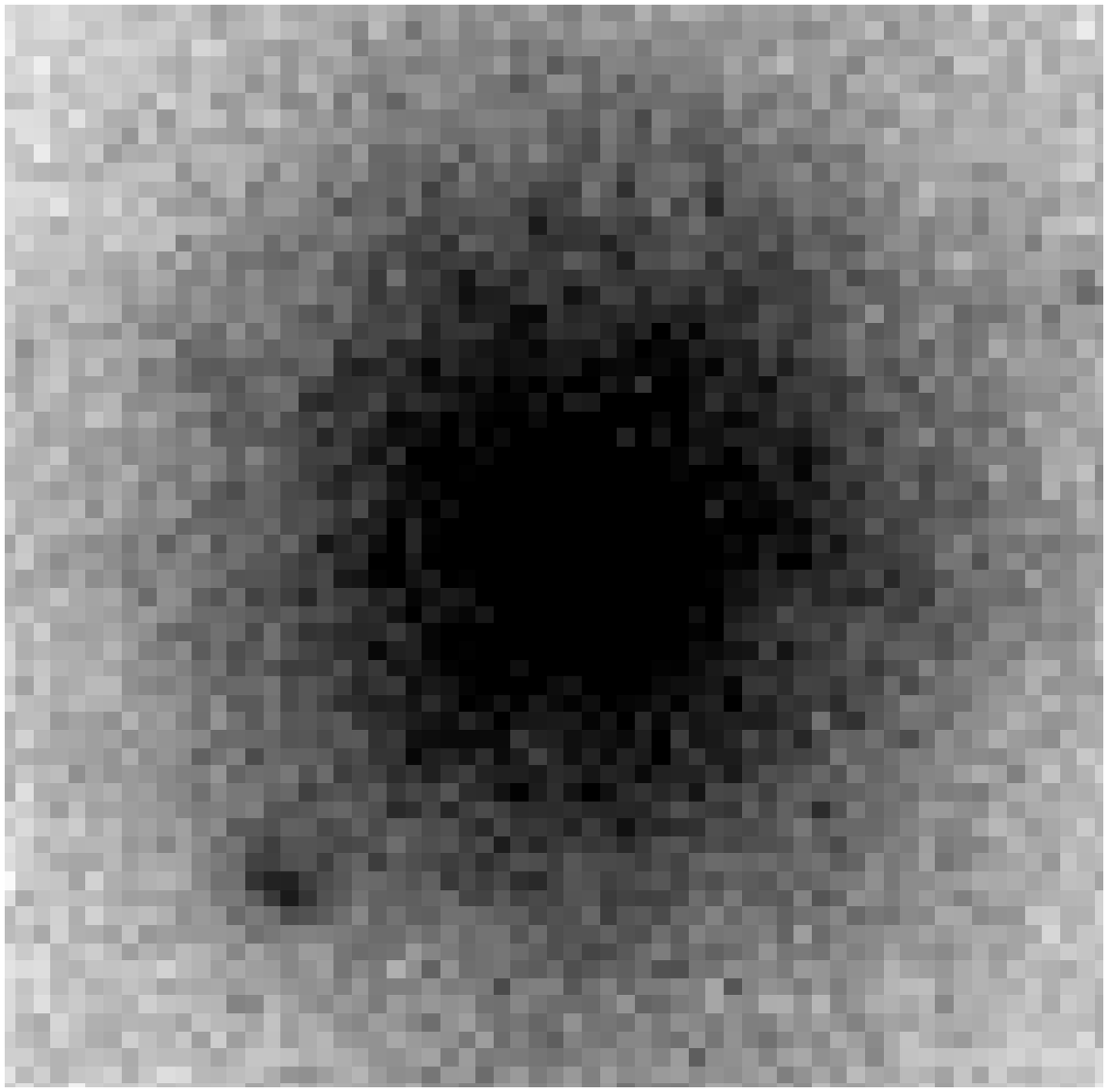,width=4.2cm}\hspace{0.2cm}\epsfig{figure=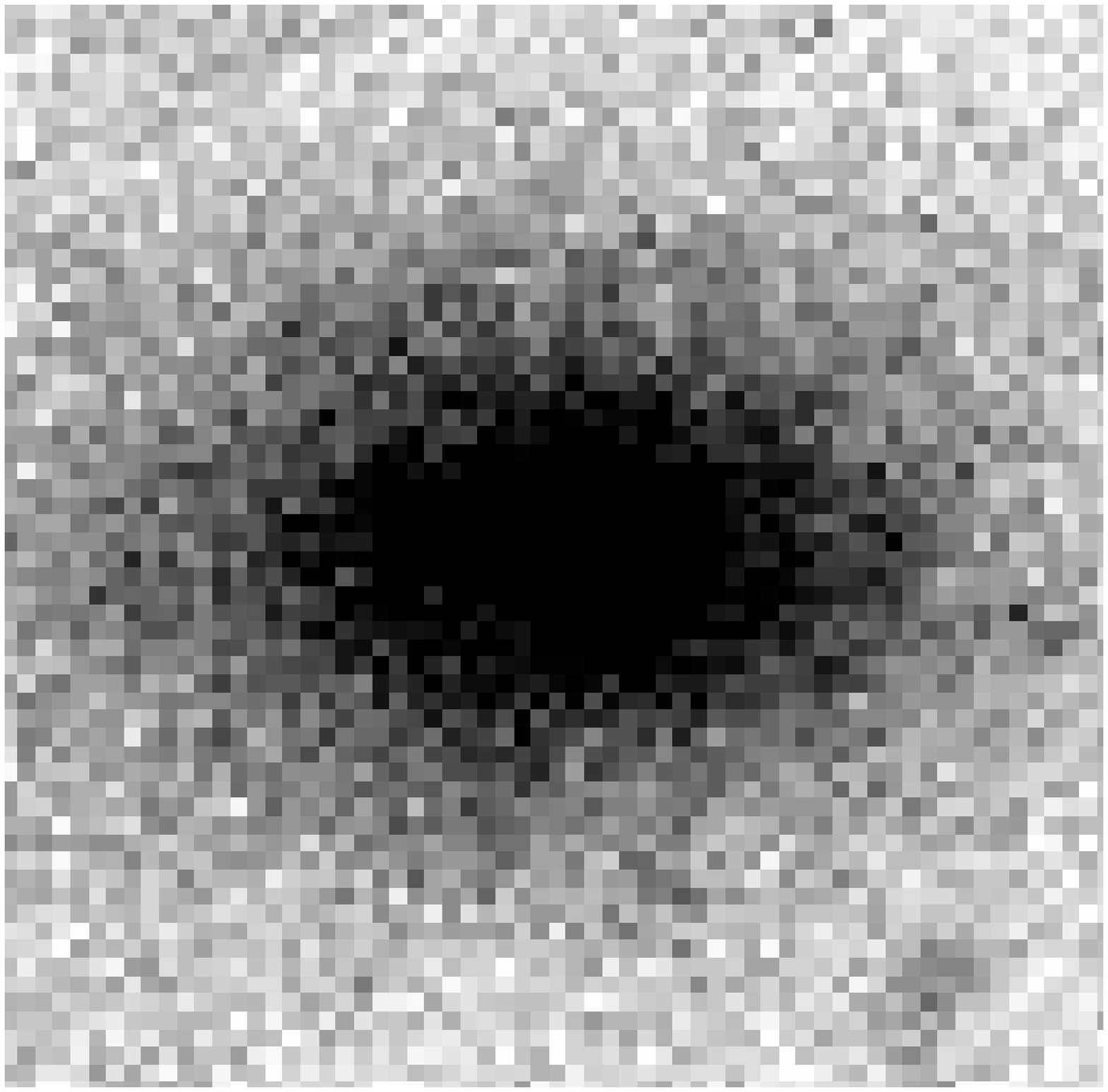,width=4.2cm}\\
\epsfig{figure=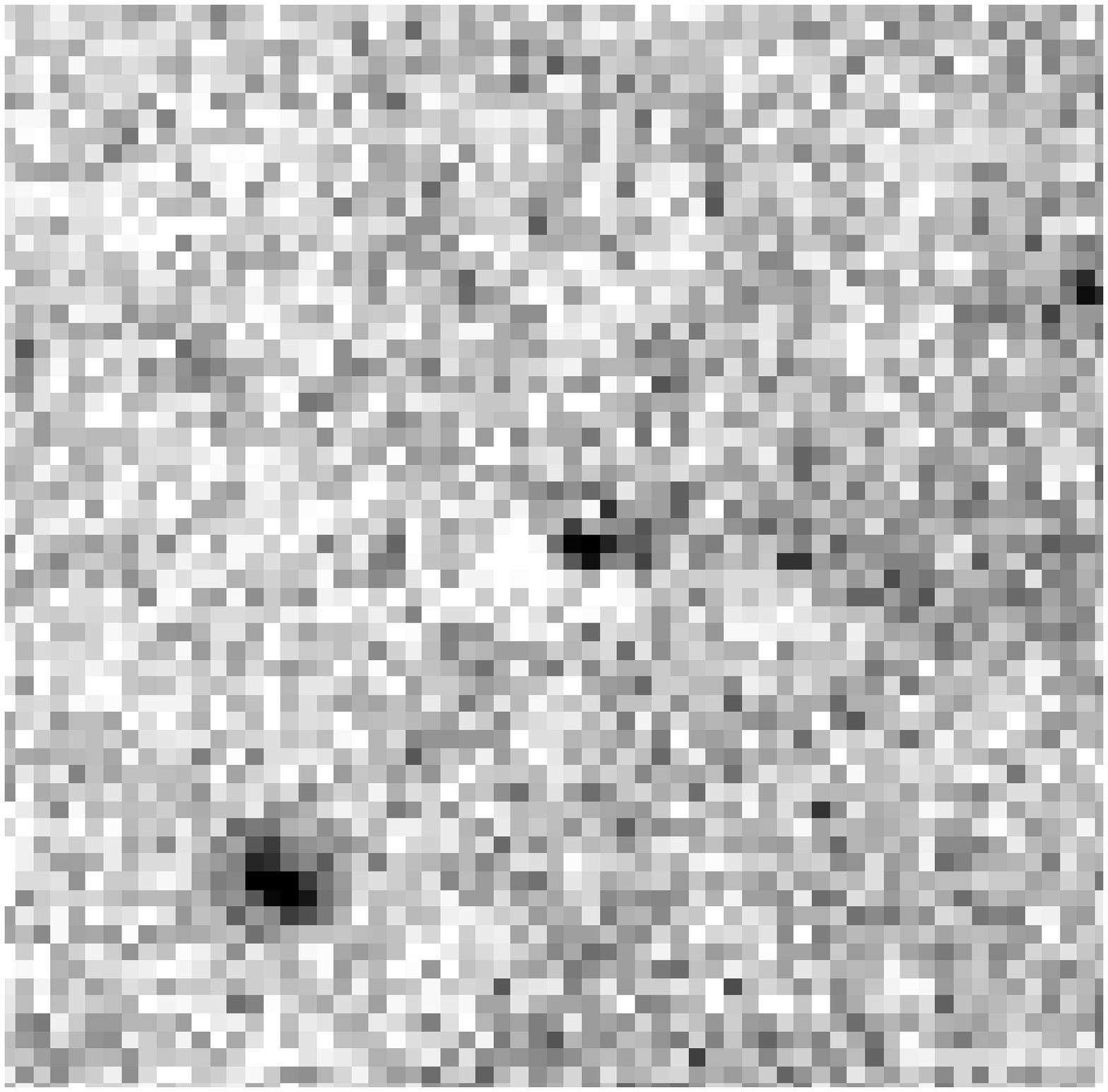,width=4.2cm}\hspace{0.2cm}\epsfig{figure=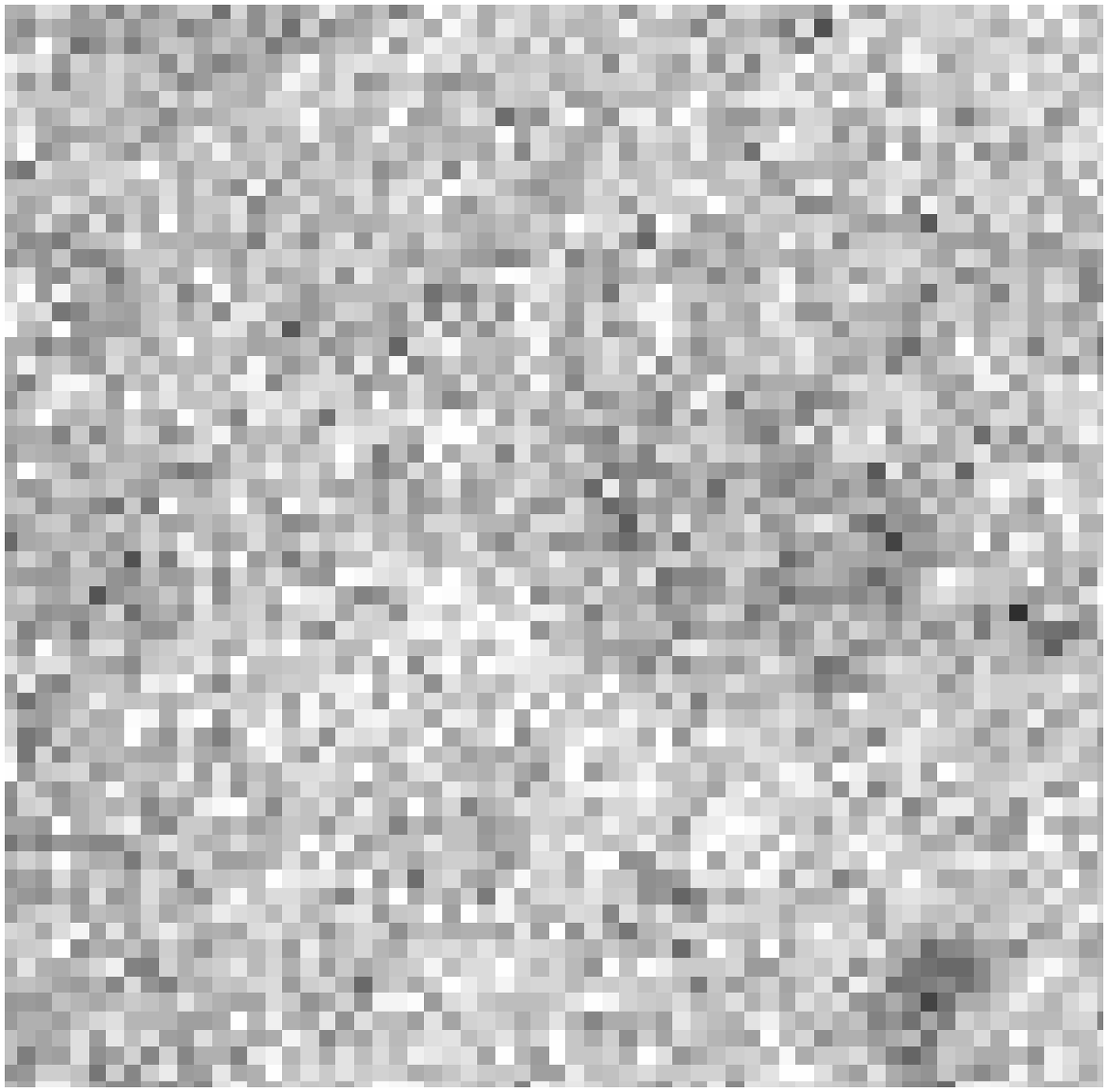,width=4.2cm}\\
\epsfig{figure=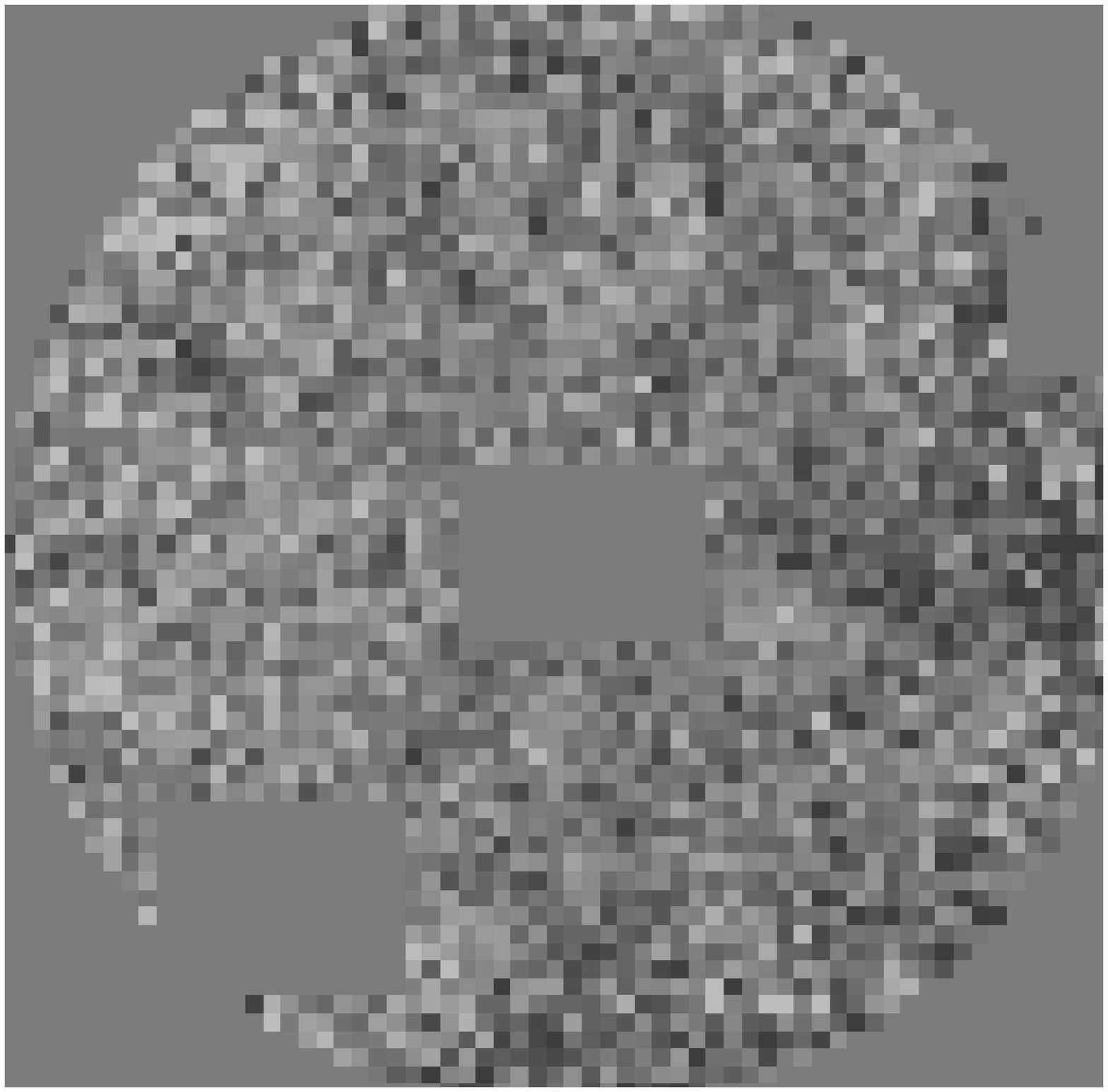,width=4.2cm}\hspace{0.2cm}\epsfig{figure=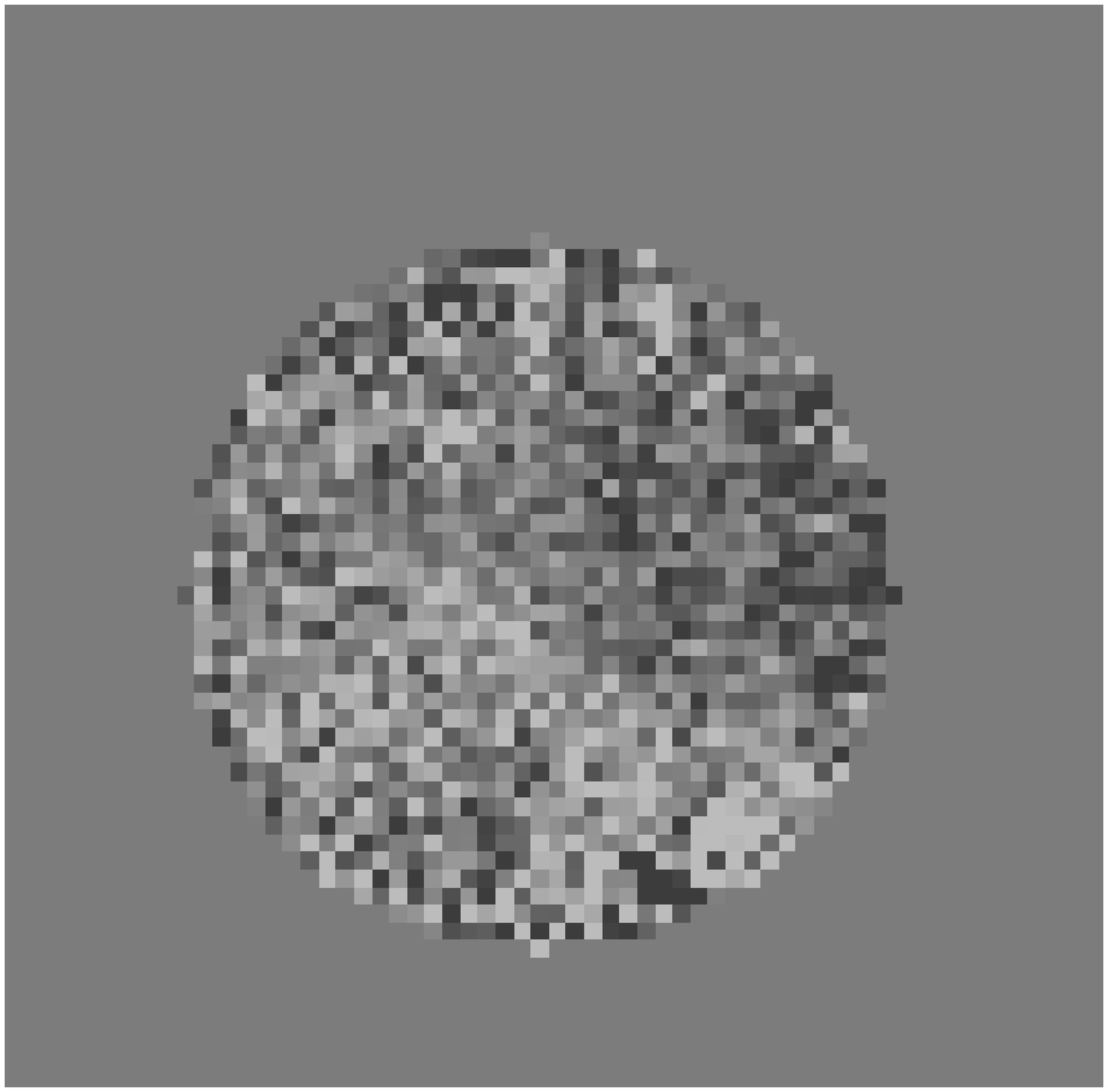,width=4.2cm}\\
\epsfig{figure=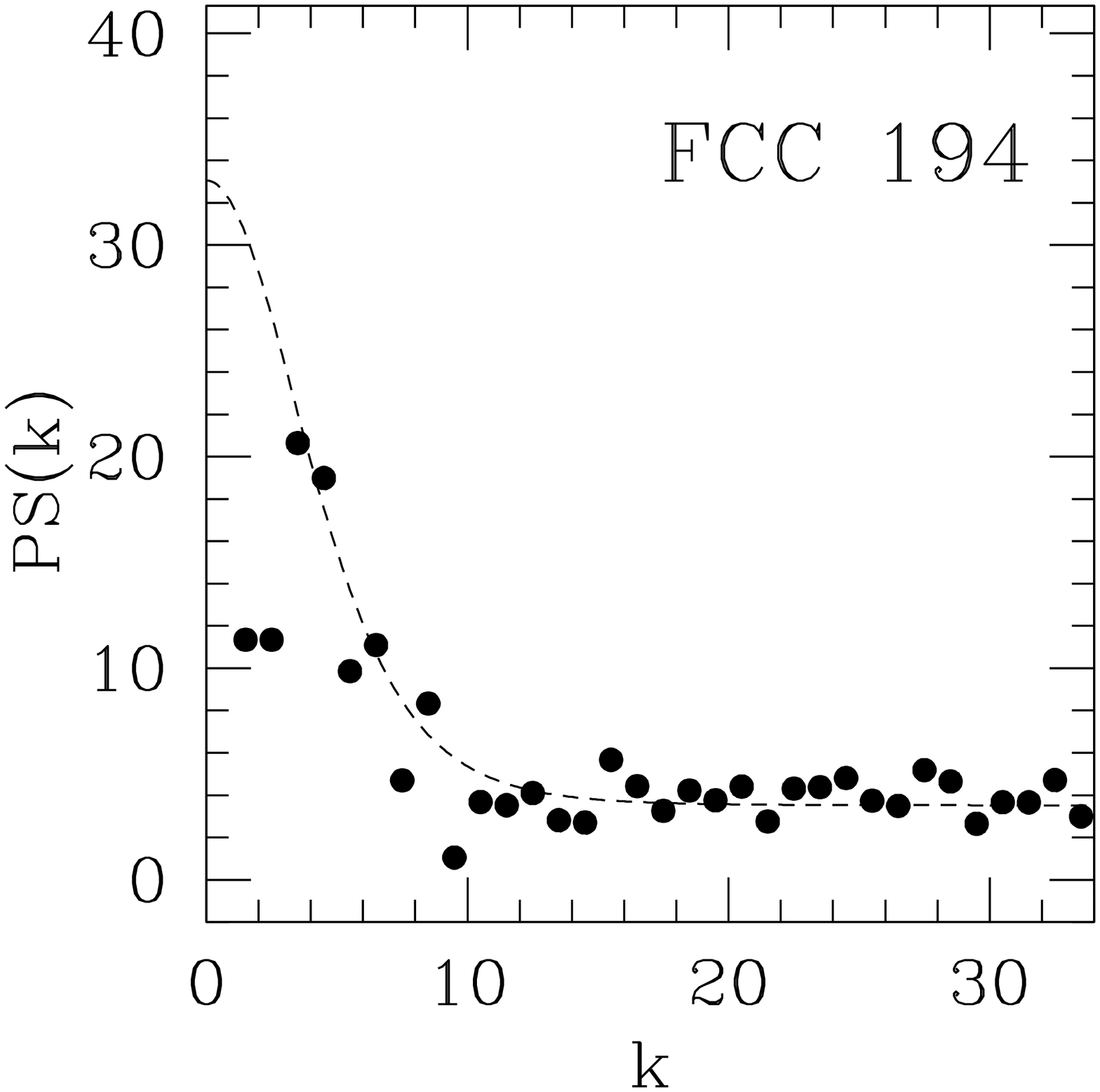,width=4.2cm}\hspace{0.2cm}\epsfig{figure=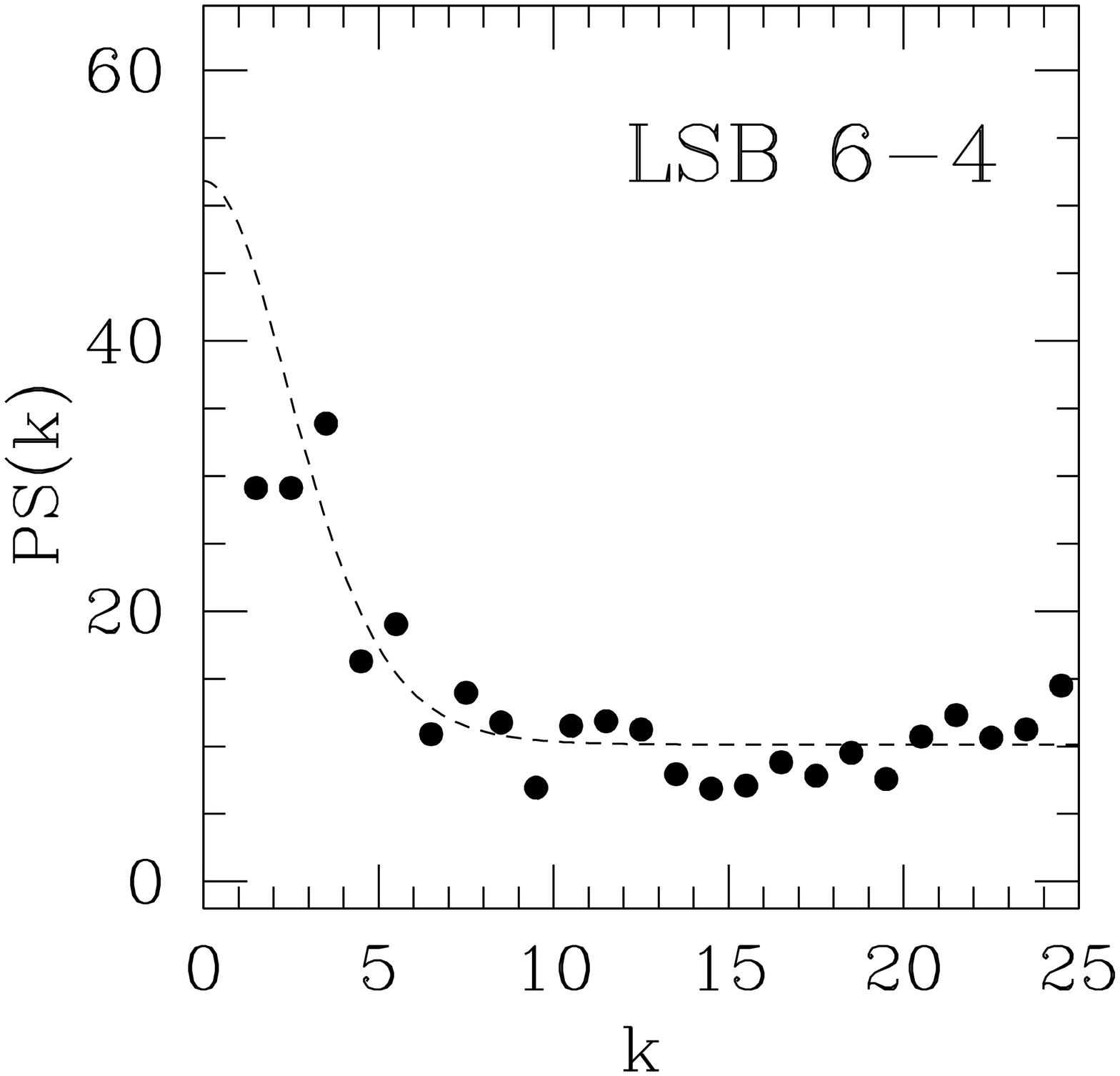,width=4.2cm}
\caption{Example thumbnails of the SBF measurement procedure for two Fornax cluster dEs. Left FCC 194, right LSB6-4. Thumbnail size is 12.5$''$. Image sequence from top to bottom: 1. Original galaxy image. 2. Galaxy light model subtracted from former image. 3. Former image divided by square root of galaxy model, with contaminating sources/fit residuals masked and with the measurement region restricted to a circular area. 4. Azimuthally averaged power spectrum of the former SBF image, with the fit indicated by the dashed line.}
\label{thumbnails}
\end{center}
\end{figure}

\begin{figure}[]
   \begin{center}
   \epsfig{figure=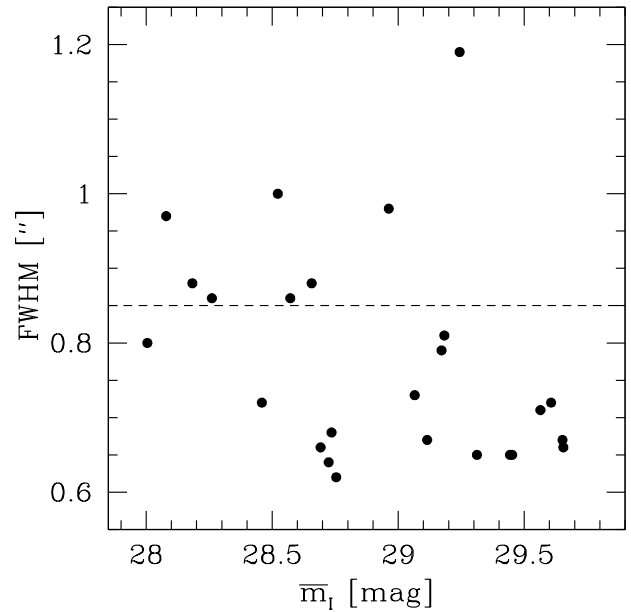,width=8.6cm}
      \caption{The fluctuation magnitude $\overline{m}_I$ plotted vs. the 
seeing FWHM of stars in the vicinity of the dEs investigated. There is a 2$\sigma$ correlation between both
entities, in the sense that bad seeing images are biased towards too bright SBF magnitudes. For the SBF calibration, we use only measurement with seeing better than 0.85$''$, as indicated by the dashed line. For those 17 galaxies, the correlation between seeing and $\overline{m}_I$ is insignificant.}
         \label{seeing}
\end{center}
   \end{figure}

A final test for observational biases before coming to the calibration is to look at the measured SBF amplitude $\overline{m}_I$ as a function of seeing, see Fig.~\ref{seeing}. Those two observables should be independent of each other. However, we do find a 2$\sigma$ significant correlation  in the sense that bad seeing data have stronger $\overline{m}_I$. We neither find a correlation between seeing and the background fluctuation amplitude $BG$ nor between seeing and galaxy luminosity.\\
One possibility to qualitatively explain the trend of seeing with $\overline{m}_I$ is that a fit to the same seeing power spectrum yields systematically higher $P_0$ when excluding more low k data points (which is the case for bad seeing data). We have tested this for our stellar power spectra: for a fixed assumed width of the power spectrum, the fitted $P_0$ depends only very marginally (to a few percent) on the limiting fit wavenumber k, even when the fitted power spectrum data points are all below 1/3 of $P_0$. We have also double-checked whether for some large seeing data we may erroneously have used marginally resolved objects for PSF fitting. However, this was not the case.\\
Jensen, Tonry \& Luppino (\cite{Jensen98}) argue that poor quality SBF data in IR SBF measurements are prone to artificial increases of the fluctuation amplitude due to sky background fluctuations, imperfect galaxy modelling or undetected background sources. For our data, we can most likely exclude the effect of sky fluctuations and undetected background sources, see Sect.~\ref{meassbffdf}. The sky fluctuations are independent of seeing and derived distance (see above), and the globular cluster fluctuations are small. We conclude that the imperfect galaxy subtraction (see e.g. right panel of Fig.~\ref{thumbnails}) is the major contributor to the increase of fluctuation power at large seeing. This is because the PSF scale approaches that expected for galaxy model residuals (the SBF images for the worst seeing data have a diameter of only about 8 PSF FHWM) such that the commonly applied rejection of low wavenumbers in the power spectrum fit can apparently not remove all the residual power for the bad seeing data.\\
For the SBF calibration, we therefore exclude all data with seeing worse than 0.85$''$ FWHM. Doing this removes the dependence of seeing on $\overline{m}_I$, see Fig.~\ref{seeing}. This exclusion applies to eight galaxies, such that we are left with a final sample of 17 SBF data points for the calibration. In the remainder of the paper the seeing restricted calibration is the reference one, but we will also indicate the calibration results for the full sample of galaxies.
\section{SBF calibration}
\label{testsbf}
\subsection{The semi-empirical approach}
\begin{figure}[]
   \begin{center}
   \epsfig{figure=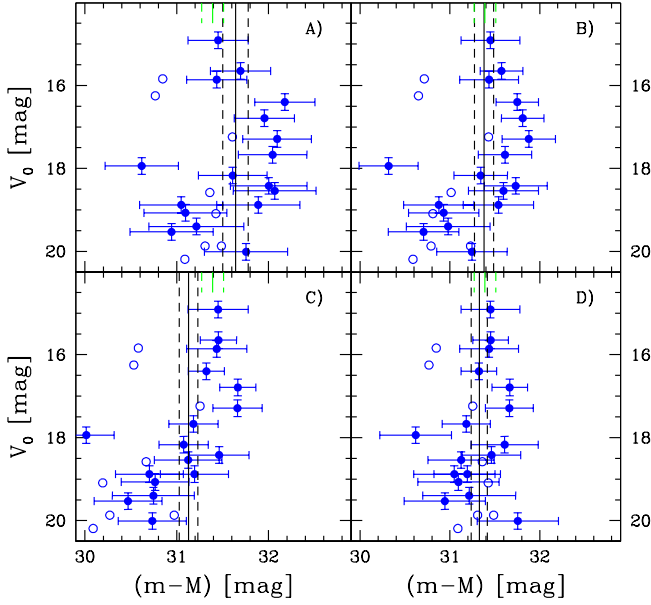,width=8.7cm}
      \caption{Apparent V magnitude plotted vs. SBF distance modulus for the four different assumed calibration relations A) to D), as defined in Sect.~\ref{sbfmodels} and Table~\ref{resultstabfdf2}. Filled circles are galaxies with seeing better than 0.85$''$. Their mean distance and its 1$\sigma$ error are indicated by the vertical solid and dashed lines. The green ticks at $(m-M)=31.39$ mag mark the reference Fornax cluster distance from Freedman et al.~(\cite{Freedm01}) and its error. Open circles are data points for galaxies with seeing worse than 0.85$''$. They were not included in calculating the mean distance and are only shown for the sake of completeness.}         \label{Vdm}
   \end{center}
   \end{figure}

\begin{figure}[]
   \begin{center}
   \epsfig{figure=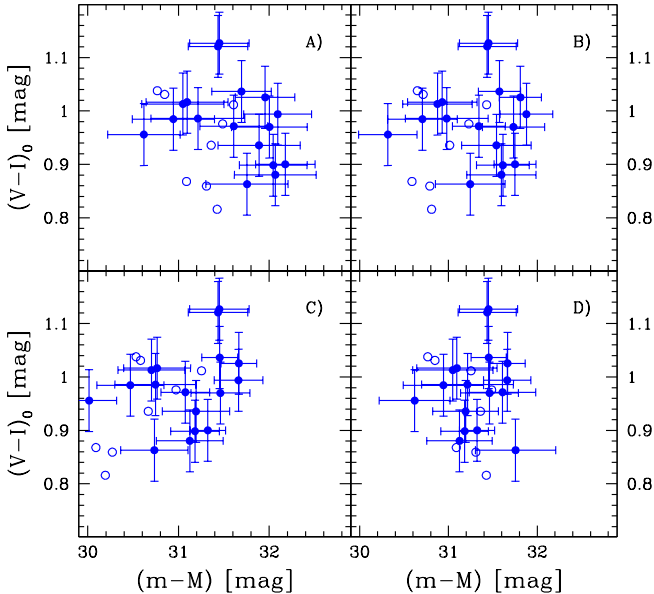,width=8.6cm}
      \caption{(V-I)$_0$ colour plotted vs. SBF distance modulus for four different assumed calibration relations, see Fig.~\ref{Vdm}.}\label{VIdm}
\end{center}
   \end{figure}

\begin{figure}[]
   \begin{center}
   \epsfig{figure=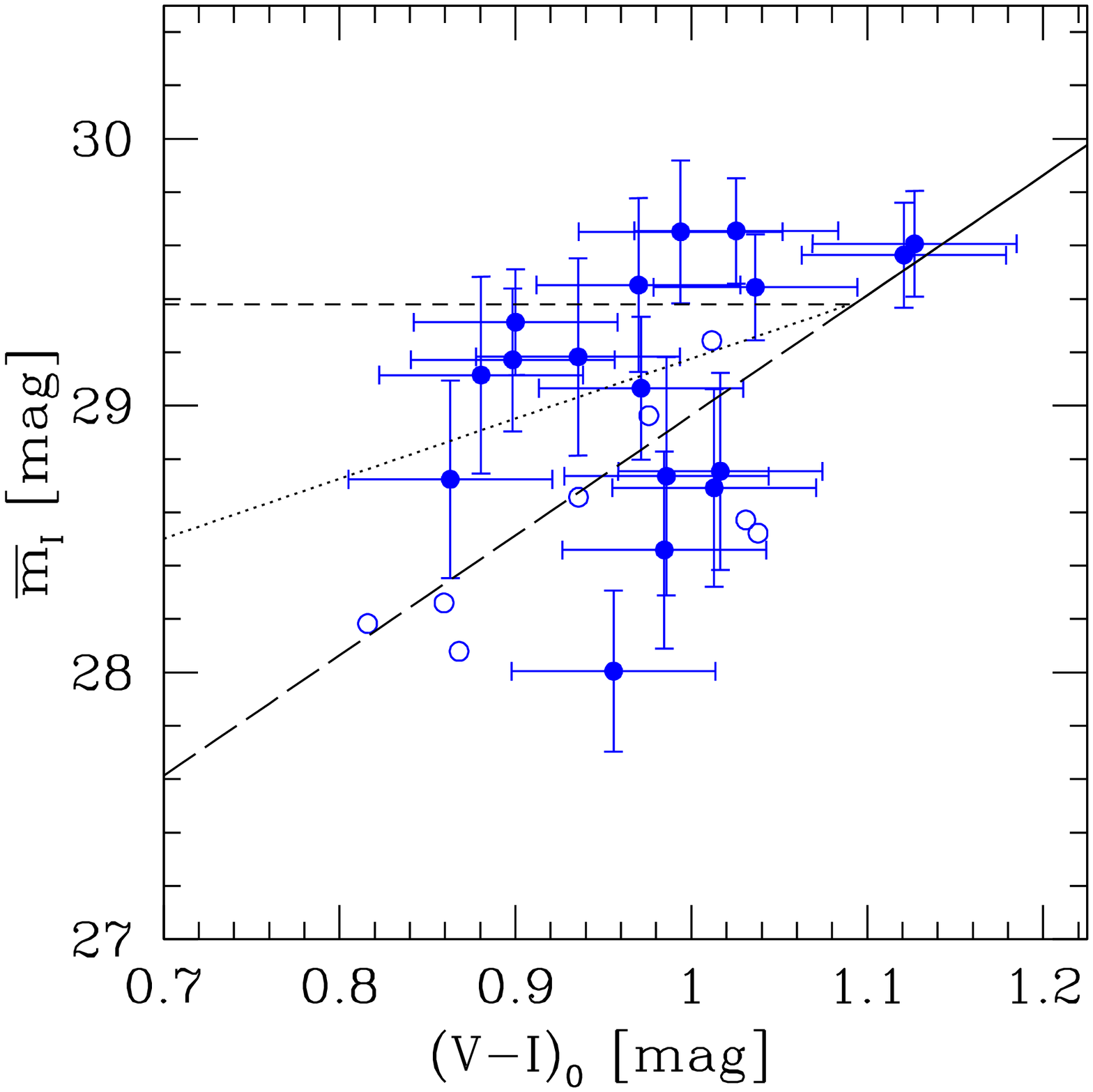,width=8.6cm}
      \caption{Apparent SBF magnitude $\overline{m}_I$ plotted vs. (V-I)$_0$ colour for the same galaxies as in Figs.~\ref{Vdm} and \ref{VIdm}. Overplotted are the calibration relations from Fig.~\ref{twobranch} for a distance modulus of $(m-M)=31.38$ mag, which is the one obtained in case B). Solid line plus dotted line is calibration case B). Case D) is defined by applying the solid and short dashed line to galaxies with $(V-I)_0<1.09$ and $\overline{m}_I$ fainter than the line for case B), while applying the solid and long dashed line to the other galaxies.}
         \label{VImbar}
\end{center}
   \end{figure}

For each of the four calibration test cases defined in Sect.~\ref{sbfmodels}, the corresponding distribution of SBF distance modulus vs. $V$ and $(V-I)_0$ is shown in Figs.~\ref{Vdm} and~\ref{VIdm}. Fig.~\ref{VImbar} plots $\overline{m}_I$ vs. $(V-I)_0$. In all plots, galaxies with seeing worse than 0.85$''$ are indicated. Table~\ref{sbfresults} gives the results in numbers.
Table~\ref{resultstabfdf2} summarises the parameters for each calibration case when using only those 17 galaxies with seeing better than 0.85$''$. Shown is the mean $(m-M)$, the distance scatter, the mean measurement error, and the dependence of $(m-M)$ on colour and magnitude ($\frac{d(m-M)}{d(V-I)_{\rm 0}}$ and $\frac{d(m-M)}{dV_{\rm 0}}$). Table~\ref{resultstabfdf3} shows the same, but for all 25 galaxies.\\
As a starting point to test these cases, we have to adopt a fiducial Fornax cluster distance modulus. The mean of the Cepheid, PNLF, GCLF and $I$-band SBF distance from the HST-Key project summary of Ferrarese et al.~(\cite{Ferrar00}) is $(m-M)=31.52$ mag. Note that here, the Cepheid scale is used to calibrate the other methods. Freedman et al.~(\cite{Freedm01}) present a revised Cepheid calibration, which causes a downward correction of 0.13 mag for the average Cepheid distance to the Fornax, Virgo and Leo I clusters (Table 8 of their paper). The resulting corrected mean Fornax distance then is 31.39 mag. This corresponds exactly to the revised Cepheid distance for Fornax from Freedman et al.~(\cite{Freedm01}), which has an error of 0.12 mag. Therefore, we define a reference distance of $(m-M)=31.39 \pm 0.12$ mag. Because of the same Cepheid scale revision, also the formal zero-point of equation~(\ref{sbfrel}) becomes fainter by 0.09 $\pm$ 0.02 mag, as calculated from comparing the revised Cepheid distances with the old ones used by Tonry et al.~(\cite{Tonry97}) for their group calibration. However, this is a small change compared to the zero-point scatter of models shown in Fig.~\ref{twobranch}. We did therefore not include this offset in the test cases, also because the Tonry calibration did not extend to blue colours.\\
We define three control criteria for the four different sets of distance moduli: 1) we demand that there is no significant correlation between colour $(V-I)_{\rm 0}$ and distance $(m-M)$. Since we aim at a calibration of $\overline{M}_I$ with $(V-I)_{\rm 0}$, this is a very fundamental requirement. 2) we require that the scatter of the distance values around their mean is not significantly smaller than the mean single measurement uncertainty. This is to avoid that we over-correct possible trends in our data. 3) we demand that there is no significant correlation between $V_{\rm 0}$ and $(m-M)$. Due to the well known colour-magnitude correlation for galaxies (see also Fig.~\ref{CMD}), this is an important double-check of criterion 1).
\begin{table*}\normalsize\small
\begin{center}
\begin{tabular}{l|rrrrr}
Calibration & $(m-M)_{SBF}$ & $\sigma_{(m-M)}$ & $\delta_{(m-M)}$ & $\frac{d(m-M)}{d(V-I)_{\rm 0}}$ & $\frac{d(m-M)}{dV{\rm 0}}$\\\hline\hline
Equ.~(\ref{sbfrel}) \bf (case A)& 31.64 $\pm$ 0.14 & 0.47$^{+0.19}_{-0.10}$ & 0.40 $\pm$ 0.02& $-$2.23 $\pm$ 1.68&$-$0.079 $\pm$ 0.079\\
Equ.~(\ref{sbfrel})~and~(\ref{sbfrelmod}) \bf (case B)& 31.38 $\pm$ 0.11 & 0.44$^{+0.18}_{-0.10}$ & 0.34 $\pm$ 0.01 & $-$0.25 $\pm$ 1.67& $-$0.135 $\pm$ 0.067\\
Constant \bf (case C)& 31.13 $\pm$ 0.10 & 0.45$^{+0.18}_{-0.10}$ & 0.31 $\pm$ 0.02& 1.775 $\pm$ 1.65& $-$0.189 $\pm$ 0.062\\
Equ.~(\ref{sbfrel})~and~constant \bf (case D)& 31.33 $\pm$ 0.09 & 0.29$^{+0.12}_{-0.06}$& 0.34 $\pm$ 0.02& 0.48 $\pm$ 1.13&$-$0.057 $\pm$ 0.050\\
\hline
&{\bf 31.39 $\pm$ 0.12}
\end{tabular}
\end{center}
\caption[]{\label{resultstabfdf2}Mean SBF distance moduli for the 17 investigated dEs with seeing better than 0.85$''$, assuming four different calibration cases (Sect.~\ref{sbfmodels}).  $\sigma_{(m-M)}$ gives the scatter of the measured distances around their mean. $\delta_{(m-M)}$ gives the mean uncertainty of a single distance measurement. The two last columns show the correlation between $(m-M)$ and $(V-I)_{\rm 0}$, and $(m-M)$ and $V_{\rm 0}$, see also Figs.~\ref{Vdm} and~\ref{VIdm}. The last line gives the Fornax cluster reference distance from the HST-Key project (Freedman et al.~\cite{Freedm01}). Note that all errors indicated are 1$\sigma$ except the error range for $\sigma_{(m-M)}$, which is a 90\% confidence range.}
\normalsize
\end{table*}
\begin{table*}\normalsize\small
\begin{center}
\begin{tabular}{l|rrrrr}
Calibration & $(m-M)_{SBF}$ & $\sigma_{(m-M)}$ & $\delta_{(m-M)}$ & $\frac{d(m-M)}{d(V-I)_{\rm 0}}$ & $\frac{d(m-M)}{dV{\rm 0}}$\\\hline\hline
Equ.~(\ref{sbfrel}) \bf (case A)& 31.52 $\pm$ 0.13 & 0.45$^{+0.19}_{-0.10}$ & 0.41 $\pm$ 0.01& $-$1.27 $\pm$ 1.14&$-$0.032 $\pm$ 0.055\\
Equ.~(\ref{sbfrel})~and~(\ref{sbfrelmod}) \bf (case B)& 31.24 $\pm$ 0.10 & 0.44$^{+0.18}_{-0.10}$ & 0.35 $\pm$ 0.02 & 0.82 $\pm$ 1.13& $-$0.102 $\pm$ 0.051\\
Constant \bf (case C)& 30.97 $\pm$ 0.09 & 0.49$^{+0.19}_{-0.11}$ & 0.32 $\pm$ 0.02& 2.92 $\pm$ 1.13& $-$0.172 $\pm$ 0.051\\
Equ.~(\ref{sbfrel})~and~constant \bf (case D)& 31.23 $\pm$ 0.08 & 0.28$^{+0.11}_{-0.06}$& 0.37 $\pm$ 0.02& 0.57 $\pm$ 0.71&$-$0.037 $\pm$ 0.033\\
\hline
&{\bf 31.39 $\pm$ 0.12}
\end{tabular}
\end{center}
\caption[]{\label{resultstabfdf3}Like Table~\ref{resultstabfdf2}, but now including all 25 investigated dEs. }
\normalsize
\end{table*}
\noindent \\Of the four test cases, the two cases B) and D) agree best with the reference distance. The null hypothesis case A) yields a marginally higher distance (1.4$\sigma$), the ``constant'' case C) a marginally lower one (1.7$\sigma$). The difference between measurement scatter and mean single measurement error is largest for case C). This is a 2.2$\sigma$ effect, consistent with non-negligible intrinsic scatter of the calibration (``cosmic variance''). As mentioned before, we are confident that our estimate of the single measurement uncertainty is not too optimistic. Therefore we take the disagreement for case C) as a hint that a constant $\overline{M}_I$ with small cosmic variance -- as predicted by the Liu and Worthey models, see Fig.~\ref{twobranch} -- does not fit our SBF measurements. The large scatter of case C) is accompanied by a clear trend of $V$ with $(m-M)$, significant at the 3$\sigma$ level. This strengthens the conclusion that case C) is not the best approximation of our data. Also in case B) there is some correlation between $V$ and $(m-M)$, but at slightly lower significance (2$\sigma$).\\
Case A) shows a marginal trend of colour with distance, but only at the 1.3$\sigma$ significance. For the shallower calibration cases B) and D) the nominal value of the trend becomes very small. For case C) the nominal slope is comparable but of opposite sign to that in case A). This is consistent with a slight flattening of the relation between $\overline{M}_I$ and $(V-I)_0$ at bluer colours.\\
Of the four test cases, case D) shows the best agreement with the data: the highest disagreement is a very marginal 1.1$\sigma$ trend of $(m-M)$ with $V$. The low distance scatter in case D) is a natural consequence of the adoption of two separate calibration branches over a common colour range.\\
We note that for the full sample, case C) is rejected at even higher significance due to a lower mean distance and stronger correlation between $(m-M)$ and $(V-I)$.
\subsection{The empirical approach}
We also give the best fit result of a purely empirical calibration valid in the colour range $0.85<(V-I)_0<1.10$ mag, assuming $(m-M)=31.39 \pm 0.12$ mag:
\begin{equation}
\overline{M}_I=-2.13 (\pm 0.17) + 2.44 (\pm 1.94) \times [(V-I)_{\rm 0} - 1.00]\;{\rm mag}
\label{sbfrelempirical}
\end{equation}
The zero point error includes our statistical errors and the error of the Fornax cluster reference distance. The statistical errors of the zero point and the slope were derived from the dispersion of the data points around the linear relation, not from the single measurement errors (this would have decreased the errors by 50\%). The formal disagreement with the Tonry slope of 4.5 is 1.1$\sigma$, while the disagreement with a 0 slope is 1.3$\sigma$. For the full sample of galaxies, the zero-point changes to $-$2.31 $\pm$ 0.15, and the slope to 2.80 $\pm$ 1.36. Not surprisingly, this agrees with the finding from the semi-empirical approach that case C) is rejected at higher significance for the full sample.\\
To quantify the intrinsic scatter -- the ``cosmic variance'' -- of the empirical ``good seeing'' calibration, we subtract in quadrature the mean single measurement uncertainty in $\overline{m}_I$ of 0.30 mag from the dispersion around the fit, which is 0.46 mag. The resulting cosmic variance is 0.34 $\pm$ 0.14 mag. This is larger than the value of 0.05-0.10 mag estimated by Tonry et al.~(\cite{Tonry97}) in their survey at redder colours, in agreement to those models that predict $\overline{M}_I$ to be more sensitive to different age/metallicity combinations in the blue than in the red (see left panel of Fig.~\ref{twobranch}). Note that the two theoretical predictions by Liu et al. and Worthey et al. in the right panel of Fig.~\ref{twobranch} that predict a colour independent $\overline{M}_I$ (case C)), do not exhibit a significant scatter.
\section{Discussion}
\label{discussion}
The results of the previous section suggest an overall shallower slope than the Tonry slope determined for red colours, accompanied by a significant cosmic scatter. The question is whether this is created by a single relation with some scatter as in case B) (see also Mei et al.~\cite{Mei05}), or by a two-branch solution as in case D) (Jerjen et al.~\cite{Jerjen01}). 
Case D) contains the implicit assumption that a) the relation between $\overline{M}_I$ and $(V-I)_0$ broadens up significantly at blue colours; b) the age-metallicity distribution of the galaxies investigated is broad; and c) the investigated galaxies consist of two distinct populations: an old, metal-poor one plus an intermediate age, more metal-rich one. Such a bimodality in stellar populations is known from the colour distribution of globular cluster systems (e.g. Peng et al.~\cite{Peng05}, Kissler-Patig~\cite{Kissle00}, Harris et al.~\cite{Harris05}, Dirsch et al.~\cite{Dirsch03}), but it has not been found for bright dEs (e.g. Geha et al.~\cite{Geha03}, Held \& Mould~\cite{Held94}). In order to support a two-branch calibration in the sense of the Jerjen papers -- as opposed to an overall shallower relation with some intrinsic scatter -- there have to be two well separated galaxy populations in the $\overline{M}_I$-$(V-I)_0$ plane.\\
To investigate this, we test the distance distribution in the different calibration cases for bimodality. We use KMM (e.g. Ashman, Bird, \& Zepf~\cite{Ashman94}) in homoscedastic mode to quantify the probability with which a double-peaked Gaussian distance distribution is preferred over a single-peaked one. This is certainly a very crude estimator, but the measurement uncertainties and low numbers do not allow for more detailed considerations. The confidence levels with which a bimodal Gaussian is preferred over a unimodal one are the following: 88\% for case A), 92\% for case B), 64\% for case C), and 27\% for case D). The strongest case for a double-peaked distribution is case B), yielding one main peak with 12 galaxies and a secondary one with five. These five galaxies are the ones below the long dashed line in Fig.~\ref{VImbar}. The fact that the probabilities for cases B) and D) are so different is consistent with a proper two-branch calibration as opposed to a unimodal one with broad scatter. However, the confidence level for case B) is only 1.8$\sigma$, and therefore this finding is still of indicative nature. We note that the SBF S/N and also the amount of background fluctuations $BG$ is not a function of distance modulus. Both entities are indistinguishable between SBF-bright and SBF-faint sample.\\
Are there any other differences between dEs belonging to the bright and faint $\overline{m}_I$ part of case D)?
The difference in mean colour $(V-I)$ between SBF-faint and bright dEs is insignificant, being -0.015 $\pm$ 0.028 mag. However, the mean $V$-band total luminosity of the SBF-faint sample is 1.6 $\pm$ 0.5 mag brighter than that of the SBF-bright sample, accompanied by a 1 mag brighter mean central surface brightness and 40\% larger exponential scale length. An obvious conclusion from this finding is that the lower luminosity and smaller dEs in Fornax have in comparison to their brighter counterparts a younger integrated age and somewhat higher metallicities, see Fig.~\ref{twobranch}. This may imply that the Fornax dwarf galaxy populations were built up in subsequent events, where the most recent events produced less massive dwarf galaxies than at earlier times. The 92\% significant bimodality in $\overline{m}_I$ is in that context consistent with at least two separate dwarf formation episodes in Fornax.\\
We finally note that the average projected radial distance to NGC 1399 is indistinguishable for both samples, indicating that both groups are not distributed in a significantly different fashion. Due to the intrinsic faintness of the investigated dEs, only three of them have spectroscopic metallicities available (Held \& Mould~\cite{Held94} and Mieske et al.~\cite{Mieske06}), of which only FCC 211 is bluer than $(V-I)_0=1.09$. This obviously does not allow one a spectroscopic confirmation of bimodality in the age-metallicity distribution.
\subsection{Revisiting the Hydra and Centaurus SBF measurements}
At the end of this discussion we revisit the relative distance between Hydra and Centaurus. The purely empirical calibration relation~(\ref{sbfrelempirical}) for $0.85<(V-I)_0<1.10$ mag can be used to check the SBF distances to the dE dominated sample of Hydra and Centaurus cluster galaxies from Mieske \& Hilker~(\cite{Mieske03b}) and Mieske et al.~(\cite{Mieske05}). There is a 0.05 mag overlap of the colour regime for relation~(\ref{sbfrelempirical}) with that of Tonry's relation~(\ref{sbfrel}). In the Hydra/Centaurus papers we assumed relation~(\ref{sbfrel}) to hold for $(V-I)_0>1.00$, and a 50\% smaller slope for $(V-I)_0<1.00$. Both the zero-point and slope of relation~(\ref{sbfrelempirical}) are consistent with that adopted for $(V-I)_0<1.00$ in those papers. That is, the SBF distances remain unchanged in the case of $(V-I)_0=1.00$ as the limit between steep and shallow slope. If we restrict the validity of relation~(\ref{sbfrel}) to Hydra/Centaurus galaxies with $(V-I)_0>1.10$ and adopt the slope of relation~(\ref{sbfrelempirical}) for bluer colours, the mean distance of {\it both} samples becomes 0.09 mag lower. I.e. there is no consequence for the relative distance. The uncertainty of relation's (\ref{sbfrelempirical}) zero-point is larger than this difference. Therefore we can state that the blue SBF calibration from this paper does not change the relative Hydra/Centaurus distance nor does it require a change in the absolute distances.
\section{Summary and conclusions}
We have presented $I$-band SBF measurements taken with IMACS@Magellan for a sample of 25 dEs in the Fornax cluster in the colour magnitude range $-16.5<M_V<-11.2$ mag and $0.8<(V-I)_0<1.10$ mag. The aim of this investigation was to address the SBF calibration at blue colours. Our colour calibration is accurate to better than 0.02 mag. To avoid observational biases towards too bright SBF amplitudes, we restrict our calibration sample to those 17 galaxies with seeing better than 0.85$''$.\\
These are our results:
\begin{itemize}
\item Our SBF data are inconsistent at the 3$\sigma$ confidence level with a constant, colour independent absolute SBF magnitude $\overline{M}_I$ at blue colours, as suggested for example in the SBF models by Liu et al.~(\cite{Liu00}) and Worthey et al.~(\cite{Worthe94}). 
\item Our SBF data support a non-zero slope between $\overline{M}_I$ and $(V-I)_0$. There is some evidence that this slope is shallower than the value of 4.5 found by Tonry et al.~(\cite{Tonry97}) at redder colours, but at less than 2$\sigma$.
\item In the $\overline{M}_I - (V-I)$ plane, our SBF data favour at the 1.8 $\sigma$ confidence level a two-branch calibration relation over a single one with some uniform scatter. We find evidence that the fainter galaxies in our samples exhibit younger integrated ages and higher metallicities than the brighter galaxies.
\item From a purely empirical SBF calibration (relation~\ref{sbfrelempirical}) we deduce a 0.34 $\pm$ 0.14 mag cosmic scatter of $\overline{M}_I$. This is several times larger than found by Tonry et al.~(\cite{Tonry97}) for redder colours, in agreement with those theoretical predictions that predict $\overline{M}_I$ to be more sensitive to different age/metallicity combinations in blue colours than at red colours.
\item Applying the empirical calibration to the SBF measurements of Hydra and Centaurus cluster dEs from Mieske \& Hilker~(\cite{Mieske03b}) and Mieske et al.~(\cite{Mieske05}) has no effect on the relative SBF distance of both galaxy clusters. Also the absolute distances do not require correction.
\end{itemize}
\label{conclusions}
\begin{acknowledgements}
We thank the anonymous referee for his comments and suggestions which helped to improve the paper. We owe thanks to the staff at Las Campanas Observatory for their assistance in carrying out the observations. SM acknowledges support by DFG project HI 855/1 and DAAD Ph.D. grant Kennziffer D/01/35298. LI was supported by FONDAP ``Center for Astrophysics''.
\end{acknowledgements}


\begin{thebibliography}{}
\bibitem[1994]{Ashman94}Ashman, K. M., Bird, C. M., Zepf, S. E. 1994, AJ, 108, 2348
\bibitem[1994]{Bertel94}Bertelli, G., Bressan, A., Chiosi, C., Fagotto, F., \& Nasi, E. 1994, A\&AS, 106, 275
\bibitem[1995]{Blakes95}Blakeslee, J.P., \& Tonry, J.L. 1995, ApJ, 442, 579
\bibitem[2001]{Blakes01}Blakeslee, J.P., Vazdekis, A., \& Ajhar, E.A. 2001, MNRAS, 320, 193
\bibitem[1997a]{Bono97a}Bono, G., Caputo, F., Cassisi, S., Castellani, V., \& Marconi, M. 1997a, ApJ, 479, 279
\bibitem[1997b]{Bono97b}Bono, G., Caputo, F., Cassisi, S., Castellani, V., \& Marconi, M. 1997b, ApJ, 489, 822
\bibitem[1991]{Bothun91}Bothun, G.D., Impey, C.D., \& Malin, D.F. 1991, ApJ, 376, 404
\bibitem[1999]{Brocat99}Brocato, E., Castellani, V., Raimondo, G., \& Romaniello, M. 1999, A\&AS, 136, 65
\bibitem[2000]{Brocat00}Brocato, E., Castellani, V., Poli, F.M., \& Raimondo, G. 2000, A\&AS, 146, 91
\bibitem[1993]{Bruzua93}Bruzual, G., \& Charlot, S. 1993, ApJ, 405, 538
\bibitem[2003]{Cantie03}Cantiello, M., Raimondo, G., Brocato, E., \& Capaccioli, M. 2003, AJ, 125, 2783
\bibitem[1991]{Castel91}Castellani, V., Chieffi, A., \& Pulone, L. 1991, ApJS, 76, 911
\bibitem[1992]{Castel92}Castellani, V., Chieffi, S., \& Straniero, O. 1992, ApJS, 78, 517
\bibitem[2003]{Dirsch03}Dirsch, B., Richtler, T., Geisler, D. et al. 2003, AJ, 125., 1908
\bibitem[1989]{Fergus89}Ferguson H.C., Sandage A., 1989, ApJ 346, 53
\bibitem[2000]{Ferrar00}Ferrarese, L., Mould, J. R., Kennicutt, R. C. et al. 2000, ApJ, 529, 745 
\bibitem[2001]{Freedm01}Freedman, W. L., Madore, B. F., Gibson, B. K. et al. 2001, ApJ, 553, 47
\bibitem[2003]{Geha03}Geha, M., Guhathakurta, P., van der Marel, R. P. 2003, AJ, 126, 1794
\bibitem[2000]{Girard00}Girardi, L., Bressan, A., Bertelli, G., \& Chiosi, C. 2000, A\&AS, 141, 371
\bibitem[1987]{Green87}Green, E.M., Demarque, P., \& King, C.R. 1987, The revised Yale isochrones and luminosity functions, New Haven: Yale Observatory, 1987
\bibitem[2005]{Harris05}Harris, W.E., Whitmore, B.C, Karakla, D., Okon, W., Baum, W.A., Hanes, D.A., Kavelaars, J.J. 2006, ApJ, 636, 90
\bibitem[1994]{Held94}Held E.V., Mould J.R., 1994, AJ 107, 1307
\bibitem[2003]{Hilker03}Hilker, M., Mieske, S., \& Infante, L. 2003, A\&AL,
397, L9
\bibitem[2005]{Hilker05}Hilker, M., Mieske, S., \& Infante, L. 2005, to appear in the proceedings of the IAUC198 "Near-Field Cosmology with Dwarf Elliptical Galaxies", H. Jerjen \& B. Binggeli (eds.), astro-ph/0505186
\bibitem[1992]{Jacoby92}Jacoby, G.H., Branch, D., Ciardullo, R. et al.,
  1992, PASP, 104, 599
  ApJ, 468, 519
\bibitem[1998]{Jensen98}Jensen, J.B., Tonry, J.L., \& Luppino, G.A. 1998, ApJ, 505, 111
\bibitem[2003]{Jensen03}Jensen, J.B., Tonry, J.L., \& Barris, B.J., 2003,
  ApJ 583, 712
\bibitem[1998]{Jerjen98}Jerjen, H., Freeman, K.C., \& Binggeli, B.  1998, AJ 116, 2873
\bibitem[2000]{Jerjen00}Jerjen, H., Freeman, K.C., \& Binggeli, B.  2000, AJ 119, 166
\bibitem[2001]{Jerjen01}Jerjen, H., Rekola, R., Takalo, L., Coleman, M., \& Valtonen, M.,
  2001, A\&A 380, 90
\bibitem[2004]{Jerjen04}Jerjen, H., Binggeli, B., \& Barazza, F.D. 2004,  AJ, 127, 771
\bibitem[2003]{Karick03}Karick, A., Drinkwater, M.J., Gregg, M.D. 2003, MNRAS, 344, 188
\bibitem[2000]{Kissle00}Kissler-Patig, M. 2000, Reviews in Modern Astronomy 13 : New Astrophysical Horizons, edited by Reinhard E. Schielicke. Hamburg, Germany : Astronomische Gesellschaft, 2000., p.13
\bibitem[2001]{Kundu01}Kundu, A., \& Whitmore, B.C. 2001, AJ, 121, 2950
\bibitem[1994]{Landolxx}Landolt, A.U. 1992, AJ, 104, 340
\bibitem[2000]{Liu00}Liu, M.C., Charlot, S., \& Graham, J. 2000, ApJ, 543, 644
\bibitem[2001]{Liu01}Liu, M.C., \& Graham, J.R. 2001, ApJ letters, 557, 31
\bibitem[2002]{Liu02}Liu, M.C., Graham, J.R., \& Charlot, S. 2002, ApJ, 564, 216
\bibitem[1993]{Luppin93}Luppino, G.A., \& Tonry, J.L. 1993, ApJ, 410, 81
\bibitem[1988]{Lynden88}Lynden-Bell, D., Faber, S.M., Burstein, D. et al. 1988, ApJ, 326, 19
\bibitem[2005]{Mei05}Mei, S., Blakeslee, J.P., Tonry, J.L. et al. (ACS VCS team) 2005, ApJ, 625, 121
\bibitem[2003a]{Mieske03a}Mieske, S., Hilker, M., \& Infante, L. 2003a, A\&A, 403, 43
\bibitem[2003b]{Mieske03b}Mieske, S. \& Hilker, M. 2003b, A\&A, 410, 455
\bibitem[2005]{Mieske05}Mieske, S.,  Hilker, M., \& Infante, L. 2005, A\&A, 438, 103
\bibitem[2006]{Mieske06}Mieske, S., Hilker, M., Infante, L. \& Jord\'an, A. 2006, AJ, 131, 2442
\bibitem[1998]{Miller98}Miller, B.W., Lotz, J., Ferguson, H.C., Stiavelli, M., \& Whitmore, B.C. 1998, ApJ, 508L, 133
\bibitem[1994]{Pahre94}Pahre, M.A., \& Mould, J.R. 1994, ApJ, 433, 567
\bibitem[2005]{Peng05}Peng, E., Jord\'an, A., C\^{o}t\'{e}, P. et al. (ACS Virgo Cluster paper IX) 2006, ApJ, 639, 95
\bibitem[1998]{Schleg98}Schlegel, D.J., Finkbeiner, D.P., \& Davis, M. 1998, ApJ, 500, 525
\bibitem[1988]{Tonry88}Tonry, J.L., \& Schneider, D.P. 1988, AJ, 96, 807
\bibitem[1997]{Tonry97}Tonry, J.L., Blakeslee, J.P., Ajhar, E.A., \& Dressler, A. 1997, ApJ, 475 399 
\bibitem[2000]{Tonry00}Tonry, J.L., Blakeslee, J.P., Ajhar, E.A., \& Dressler, A. 2000, ApJ, 530, 625
\bibitem[2001]{Tonry01}Tonry, J.L., Dressler, A., Blakeslee, J.P. et
  al. 2001, ApJ, 546, 681 
\bibitem[1985]{Vanden85}Vandenberg, D.A. 1985, ApJS, 58, 711
\bibitem[1996]{Vazdek96}Vazdekis, A., Casuso, E., Peletier, R.F., \& Beckman, J.E. 1996, ApJS, 106, 307
\bibitem[1994]{Worthe94}Worthey, G. 1994, ApJS, 95, 107
\end{thebibliography}
\end{document}